\documentclass[a4paper,fleqn,usenatbib]{mnras}

\usepackage[T1]{fontenc}
\usepackage{ae,aecompl}
\usepackage{widetext}

\usepackage{graphicx}	
\usepackage{amsmath}	
\usepackage{amssymb}	
\usepackage{hyperref}      

\def\refnew#1{\,(\ref{#1})}

\newcommand{\appropto}{\mathrel{\vcenter{
  \offinterlineskip\halign{\hfil$##$\cr
    \propto\cr\noalign{\kern2pt}\sim\cr\noalign{\kern-2pt}}}}}

\def\Enode{E_{\mathrm{node}}}
\def\vcv{v_{\mathrm{cv}}}

\def\OS{\Omega_S} 
\def\norb{\Omega_{\mathrm{orb}}} 
\def\Tosc{T_{\mathrm{osc}}} 
\def\Phifirst{\Phi_\mathrm{1}} 
\def\rb{r_b} 

\title[]{How {\it Cassini} Can Constrain Tidal Dissipation in Saturn}

\author[Jing Luan, et al.]{
Jing Luan,$^{1}$\thanks{E-mail: jingluan@berkeley.edu}
Jim Fuller,$^{2,3}$
Eliot Quataert$^{1}$
\\
$^{1}$Astronomy Department, University of California at Berkeley, Berkeley, CA 94720, US\\
$^{2}$Kavli Institute for Theoretical Physics, Kohn Hall, University of California, Santa Barbara, CA 93106, USA\\
$^{3}$TAPIR, Walter Burke Institute for Theoretical Physics, Mailcode 350-17, Caltech, Pasadena, CA 91125, USA
}

\date{Accepted XXX. Received YYY; in original form ZZZ}

\pubyear{2017}

\begin{document}
\label{firstpage}
\pagerange{\pageref{firstpage}--\pageref{lastpage}}
\maketitle

\begin{abstract}
Tidal dissipation inside giant planets is important for the orbital evolution of their natural satellites. It is conventionally treated by parameterized equilibrium tidal theory, in which the tidal torque declines rapidly with distance, and orbital expansion was faster in the past. However, some Saturnian satellites are currently migrating outward faster than predicted by equilibrium tidal theory. Resonance locking between satellites and internal oscillations of Saturn naturally matches the observed migration rates. Here, we show that the resonance locking theory predicts dynamical tidal perturbations to Saturn's gravitational field in addition to those produced by equilibrium tidal bulges. We show that these perturbations can likely be detected during {\it Cassini}'s proximal orbits if migration of satellites results from resonant gravity modes, but will likely be undetectable if migration results from inertial wave attractors or dissipation of the equilibrium tide. Additionally, we show that the detection of gravity modes would place constraints on the size of the hypothetical stably stratified region in Saturn. 
\end{abstract}




\section{Introduction}
Some of the Saturnian satellites are believed to have migrated outward due to tidal interaction with Saturn \citep[][ and reference therein]{Peale}.
The tidal origin of their migration was first proposed by \cite{Goldreich-1965}. 
Tidal interaction is conventionally treated assuming that equilibrium tides are dissipated by a fraction, $1/Q$, every cycle\footnote{A cycle could be one orbit period of satellite or one rotation period of planet or some combination of them, depending the specific case.}. This approximation originates from studies of terrestrial bodies in the solar system \citep[e.g.][]{Jeffreys}. The frequencies of free oscillations of small solid bodies are much higher than that of tidal forcing in most cases, and therefore their instantaneous tidal deformation is well approximated by the equilibrium tide. However, caution should be taken when extending this theory to gaseous planets, because they are larger and less dense and thus they may support free oscillation modes or waves whose frequencies match those of tidal forcing. The conventional treatment, i.e., dissipation of the equilibrium tide with $Q$ independent of time, is so convenient that it is widely applied to both planets and stars. The resulting tidal torque decreases by the sixth power of distance \citep[e.g.][]{Murray-Dermott}. Consequently, a conceptual belief has been established over years that tidal interaction weakens rapidly with distance, and that most of the satellite's orbital expansion took place in the distant past when the satellites were closer to Saturn.

However, \citet{Lainey} recently reported surprisingly fast ongoing migration for Enceladus, Tethys, Dione, and Rhea. This observation disfavors the conventional belief of equilibrium tidal dissipation. Equilibrium tides, as long as getting damped by a constant efficiency\footnote{\cite{Shoji} propose that damping in a viscoelastic core of Saturn may have a damping efficiency varying with frequency.}, face a common problem, i.e. they require the satellites to be much younger than the solar system, since they yields tidal torques decaying steeply with distance\footnote{Late formation of Saturn's satellites has been proposed \citep{Charnoz,Cuk}.}. Instead, \citet{Fuller-2016} propose that satellites enter resonance locks with internal oscillations of Saturn. The planet's oscillation frequencies and satellite orbital frequencies evolve together such that resonances can be maintained over long time scales. \citet{Fuller-2016} predict orbital migration rates consistent with observations, assuming that the oscillation frequencies evolve on the thermal timescale of Saturn. In this scenario, the migration of satellites is controlled by the evolution of Saturn's interior, which is independent of the distance of the satellites from Saturn. The satellite distance does affect capture or breaking of resonance locks, which will be described qualitatively below but is not the focus of this paper.

Resonance locking, first introduced into astronomy by \citet{Witte-Savonije}, is similar to surfing. A surfer slides sideways on a wavefront, gaining just the right amount of speed to move together with it. A surfer and ocean wave are analogous to a satellite and Saturnian oscillation. The essential difference is that a satellite excites the oscillation of Saturn, whereas an ocean wave propagates independently of a surfer. It seems to violate energy conservation that a satellite gains energy and angular momentum from an oscillation it excites, but it does not. Because Saturn rotates faster than a satellite orbit, the oscillation of Saturn propagates prograde with the satellite in the inertial frame, but retrograde relative to the frame co-rotating with Saturn at angular frequency $\OS$. In the inertial frame, Saturn contains less energy and angular momentum in the presence of the oscillation than in its absence \citep{Pierce}. Therefore, the oscillation excited by the satellite contains negative energy and angular momentum, whereas the satellite gains positive energy and angular momentum that originates from the rotation of Saturn, and the oscillation of Saturn is merely an intermediary. 

However, capture into a resonance lock is not guaranteed. Consider the case in which the interior evolution of Saturn pushes an oscillation towards resonance with a satellite. The oscillation gets excited by the tidal force of the satellite, but it also gets damped through dissipative processes, e.g., heat diffusion and turbulent viscosity. The damping produces a phase lag between the satellite and the oscillation of Saturn, leading to a positive torque on the satellite. The tidal torque is proportional to the phase lag and the energy of the oscillation. The former is $\propto \gamma$, the damping rate of oscillation, and the latter is $\propto A^2$, where $A$ is the amplitude of the oscillation. The tidal torque, $T_{\mathrm{osc}}\propto \gamma A^2$, grows near resonance\footnote{The growth rate of the mode energy, $d(A^2)/dt$, also contributes to $T_{\mathrm{osc}}$. It needs to be taken into account if we consider the capture probability of the resonance lock, which is not the topic of this paper. Here, we assume the system is in resonance lock. It is in an equilibrium state (stable fixed point) and $A$ hardly changes.}. 
The resonance lock will succeed if the tidal torque becomes large enough for the satellite to evolve at the same rate as the oscillation in the frequency domain. Otherwise, the oscillation sweeps past the satellite in the frequency domain, torquing it temporarily, but failing to lock it. 

The tidal torque can grow near resonance through two mechanisms.  A resonance with a gravity mode (g-mode) increases $A$ but keeps $\gamma$ constant. G-modes can exist if stable stratification is present inside Saturn. An inertial wave attractor, in contrast, increases $\gamma$ but keeps $A$ constant. Hence, to produce the same torque on a satellite, the two mechanisms perturb the external gravity field differently.  The potential perturbation has dependence $\Phi^\prime\propto A$ but is almost independent of $\gamma$\,\footnote{The dissipation rate, $\gamma$, determines the phase lag of the oscillation pattern and therefore also determines the phase of $\Phi^\prime$. But the phase lag itself is probably too tiny to measure.}. Therefore, a resonant gravity mode or inertial wave attractor are distinguishable from a gravity measurement. The {\it Cassini} spacecraft, currently in its proximal orbits, will fly by the surface of Saturn $22$ times by September 2017 \citep{Dunford}. Ten of those close encounters are dedicated to measure the gravitational field of Saturn\,\footnote{{\it Cassini} sends radio signals at certain wavelengths back to earth. Its velocity along the line of sight is measured through Doppler shifts. Its acceleration is then extracted from the velocity as a function of time.}. The anticipated accuracy is unprecedented,\footnote{Through private communication with Phillip D. Nicholson and Luciano Iess.} and may allow us to  distinguish between a resonant g mode and an inertial wave attractor, or at least constrain their parameters. 

Mechanisms proposed to damp equilibrium tides, including turbulent viscosity \citep{Goldreich-Nicholson,Zahn-1966}, viscoelastic core \citep{Remus,Guenel} and elliptical instability \citep{Kerswell,Cebron}, may also damp a resonant g mode or inertial wave. Due to the uncertainty of the damping mechanism, we treat the damping rate, $\gamma$, as a free parameter in the main text. This paper focuses on possible observational signatures to reveal the suggested resonance locking. The specific damping mechanism is thus a secondary point\footnote{ Without knowing which internal oscillation mode or wave is resonantly locking which satellite, it is not even practical to estimate $\gamma$ according to a specific mechanism.}.

This paper is arranged as follows. Section~\ref{sec:candidates} describes how a g mode and an inertial wave attractor work in a resonance lock. Section~\ref{sec:MMR} comments on the influence of orbital mean motion resonances (MMRs) on resonance locks, because most of Saturn's major satellites are involved in MMRs. We estimate perturbations to the gravitational potential of Saturn by a g mode and an inertial wave attractor in Section~\ref{sec:gravity-anomaly}. Section~\ref{sec:compare} compares our results with the expected accuracy of the gravity measurement by {\it Cassini}. Section~\ref{sec:f-mode-main-body} estimates the gravitational potential due to the fundamental modes of Saturn which are proposed to excite the observed density waves in the C ring \citep{Hedman-Nicholson}. We find it far below the anticipated detection threshold of {\it Cassini}. In Section~\ref{sec:conclusion}, we discuss our results and present our conclusions. Readers mainly interested in indications for observation are suggested to read through Section~\ref{sec:candidates} to get the basic idea of resonance locking and then focus on Section~\ref{sec:compare}. 

\section{Candidates for resonance lock}\label{sec:candidates}
In this section, we discuss two different ways resonance locking operates, assuming successful locking in each case. 

\subsection{Gravity modes}\label{sec:g-mode}
A gravity mode propagates only in stably stratified regions in which buoyancy is able to restore oscillations \citep[e.g.][]{Cox}. Seismology of Saturn's ring system reveals fundamental modes
 in Saturn \citep{Hedman-Nicholson}. Fine splitting of those fundamental modes indicates the existence of stable stratification inside Saturn \citep{Fuller-2014}. 

The amplitude of a gravity mode grows as its frequency converges with the tidal forcing frequency of a satellite. It is instructive to visualize the propagation cavity of a g mode as a spring, and tidal force of the satellite swings the outer end of the `spring'.\footnote{The tidal force of the satellite operates everywhere, but it is strongest at the outer end of the propagation cavity, because tidal gravity potential is $\propto (r/a)^{l}$ \citep[e.g.][]{Murray-Dermott}.} The excited oscillatory motion then propagates inward, which is essentially an ingoing g wave. It gets reflected at the inner boundary of the propagation cavity, and returns to the outer boundary, forming an outgoing g wave. Ingoing and outgoing g waves with the right relative phase compose a standing g wave, i.e., a g mode. Consider a g mode with $n_r$ radial nodes and angular frequency $\sigma_g$. The phase of the g wave increases by $2\pi n_r$ as it returns to the outer boundary after one reflection. Meanwhile, the tidal force changes its phase by $2\pi n_r \sigma/\sigma_g = 2\pi n_r (1+\delta \sigma/\sigma_g)$, where $\sigma=\sigma_g+\delta\sigma$ is the angular frequency of the tidal forcing. As long as the frequency mismatch, $|\delta\sigma|\ll \sigma$, the wave remains nearly in phase with tidal force, so its amplitude increases. The amplitude grows by roughly the same amount every time the wave returns to the outer boundary. After $\sim 1/|\delta \sigma|$, the wave shifts out of phase with respect to the tidal force, and its amplitude saturates. The saturation amplitude is $A_{\mathrm{sat}}\propto 1/|\delta\sigma|$.

As a g mode frequency converges with the tidal forcing frequency, $|\delta\sigma|$ decreases. In the case of slow convergence, which applies to Saturn, a g mode has enough time to reach $A_{\mathrm{sat}}$ at any given instant. Therefore, the amplitude of a g mode, $A\sim A_{\mathrm{sat}}\propto 1/|\delta\sigma|$, increases upon resonance. The essential reason is that the g mode has a well defined propagation cavity such that it returns to where it gets excited after having accumulated an integer multiple of $2\pi$ in phase. We will see that inertial waves do not share this property and therefore their amplitude does not grow upon resonance. 

On the other hand, the damping rate, $\gamma$, remains the same as long as a g mode stays within the linear regime, i.e. $\xi_r k_r\ll 1$. Turbulent viscosity in convective regions and heat diffusion in stably stratified regions both damp g modes. They are estimated in Appendix~\ref{sec:gamma-g-mode} to be
\begin{equation}\label{eq:gamma-turb-sim}
\gamma_{\mathrm{turb}}\sim  {0.1\sim 1\over (n_r+1)} \,\mathrm{Gy}^{-1}\, ,
\end{equation}
and 
\begin{equation}\label{eq:gamma-diff-sim}
\gamma_{\mathrm{diff}}\sim {(0.1\sim 1)}(n_r+1)\,\mathrm{Gy}^{-1}\, ,
\end{equation}
which are both small. Turbulent viscosity is weak because convection in Saturn turns over on a timescale much longer than the typical oscillation period, $\sim \OS^{-1}$. Therefore eddies as large as the local scale height do not act like viscosity \citep{Goldreich-Nicholson}. Eddies turning over on a timescale similar to or shorter than the oscillation period are downward in the turbulent cascade. They are small and slow, and for a Kolmogorov cascade, they have velocity, $v\propto l^{1/3}$, where $l$ here refers to the linear size of eddy. Turbulent viscosity is weak because kinetic viscosity is roughly the length multiplied by the velocity of the eddy. Damping by heat diffusion is weak as well, mainly because the current thermal timescale of Saturn is long. Note that $\gamma_{\mathrm{diff}}$ is independent of the specific mechanism for heat diffusion, as demonstrated in Appendix~\ref{sec:gamma-diffusion}. It could be created by diffusion through radiation or conductivity. 

We acknowledge that $\gamma$ for g modes is very uncertain. There may exist other damping mechanisms beyond our knowledge, e.g. damping through conversion to inertial waves, which we briefly discuss in Section~\ref{sec:compare}. Fortunately, as we will see in Section~\ref{sec:compare}, our main results depend on $\gamma$ weakly.


\subsection{Inertial wave attractors}\label{sec:inertial-wave-attractor} 

Inertial waves are restored by the Coriolis force, $-2\bmath{\OS}\times \dot{\bmath{\xi}}$, and therefore they reside in rotating bodies within the frequency range $-2\OS<\sigma<2\OS$ \citep{Greenspan}.\footnote{For tidally excited oscillations this condition is satisfied as long as the azimuthal order of the oscillation mode, $m$, is less than or equal to two.} The WKB dispersion relation is
\begin{equation}
\sigma={|2\mathbf{\OS}\cdot \mathbf{k}| \over k}\, ,
\end{equation}
i.e., the angle $\beta$ between the spin axis and the wave vector satisfies $\cos\beta=\pm \sigma/(2\OS)$. Therefore, reflection of inertial wave rays is nonspecular except when the reflection plane is perpendicular to the spin axis. Nonspecular reflection prevents inertial waves from returning to where they are excited, and thus, unlike g modes, the amplitudes of inertial waves do not grow. This heuristic is not exact but serves an intuitive way for understanding inertial waves. Strict mathematical development is found in \cite{Ogilvie-2013}, which we will briefly review in Section~\ref{sec:compare}.

Inertial waves do not form standing waves as normally defined \citep{Greenspan}, and therefore they are usually not referred to as inertial modes. However, at certain frequencies, after multiple reflections inertial wave rays converge toward a spatial pattern  called a wave attractor \citep[e.g.][]{Maas}. An inertial wave attractor closes in space, and therefore is analogous to a mode. However, a mode is identified by quantum numbers, namely the numbers of radial and angular nodes, whereas wave attractors may not be quantized in a similar way.

Inertial wave attractors usually form at discrete frequencies, at which the damping of inertial waves peaks \citep[e.g.][]{Ogilvie-Lin}. \cite{Ogilvie-2013} show that smaller kinetic viscosity sharpens wave attractors, making their peaks in tidal dissipation narrower and higher, until nonlinear damping starts to operate. Linear damping scales with the square of the velocity shear multiplied by kinetic viscosity. Nonlinear damping, e.g., shock breaking or generation of turbulence, also contributes to $\gamma$ if the velocity shear exceeds the linear regime. Either way, inertial wave attractors promote damping, i.e., $\gamma$ increases as a satellite's tidal forcing frequency approaches the frequency of an inertial wave attractor.

Inertial wave attractors form by multiple reflections. Reflection, or more generally speaking, scattering conserves the total action of a wave, which is the classical physics analogue to the number of quanta in quantum physics. The total energy of an inertial wave is proportional to the action multiplied by $\sigma$. It follows that $A$ is conserved by reflection, since the total wave energy is $\propto A^2$. Therefore, the formation of inertial wave attractors does not change $A$. 

Since the tidal torque scales as $\Tosc \propto \gamma A^2$, the torque on a satellite increases as it approaches an inertial wave attractor in the frequency domain. However, the physical mechanism underlying resonance locking differs between an inertial wave attractor and a g mode. To summarize, a g mode increases $A$ while keeping $\gamma$ constant, whereas an inertial wave attractor increases $\gamma$ while keeping $A$ constant.

\section{Mean motion resonance}\label{sec:MMR}

Saturn has three pairs of satellites involved in orbital MMRs \citep[e.g.][]{Murray-Dermott}, in which mutual interaction fixes the orbital period ratios of the satellites. The MMRs are Mimas-Tethys in a $4:2$ inclination MMR, Enceladus-Dione in a $2:1$ eccentricity MMR, and Titan-Hyperion in a $5:3$ eccentricity MMR \citep{Urban}. Both satellites in each pair must share the same long-term migration rate $\langle \dot a/a\rangle$, i.e., the migration rate averaged over billion year timescales.

\citet{Lainey} report a migration rate, $\dot a/a \sim 1/(5\,\mathrm{Gyr})$, for Enceladus, Tethys, Dione and Rhea. Data sets with time spans of $\sim 100\,\mathrm{yr}$ and $\sim 20\,\mathrm{yr}$ are analyzed independently, producing consistent results. Mimas had been reported to migrate inward by \citet{Lainey-2012}. However, it should migrate outward together with its MMR companion, Tethys. \citet{Luan} speculate that the inward migration of Mimas may be biased by the MMR torque which overwhelms the migration torque by a factor of $\sim 10^5$ and librates every $\sim 80\,\mathrm{yr}$. The data set spanning over $\sim 100\,\mathrm{yr}$ employed by \citet{Lainey-2012} most likely does not completely average out the MMR torque. There are not yet published migration rates for Titan or Hyperion. 

Convergent migration is necessary for MMR capture. Assuming both satellites migrate due to resonance locking before they get caught in a MMR, the corresponding oscillations of Saturn must evolve convergently in the frequency domain. However, this requirement is not naturally satisfied by resonance locking, but instead depends on the evolution of the interior of Saturn, of which we lack enough understanding to accurately assess. 

The outer satellite in a MMR is likely no longer in a resonance lock with Saturn. Once captured in a MMR, the satellite migrates in the frequency domain at the same rate as the inner satellite, rather than the oscillation of Saturn locking it in the past. Therefore, a resonance lock is broken by the formation of a MMR. It follows that Tethys, Dione and Hyperion are not currently in a resonance lock with an oscillation of Saturn. In addition, to maintain a MMR, the inner satellite must provide the outer satellite with angular momentum, which originates from the tidal torque by a resonance lock with Saturn. Therefore, a MMR must increase the amplitude of a resonantly locked g mode, or the damping rate of a resonant inertial mode attractor. 

An oscillation of Saturn, and the inner and outer satellites form a resonance chain. The planet and inner satellite are linked by a resonance lock, while the inner and outer satellites are linked by a MMR. All three must evolve together in the frequency domain. Hence, an oscillation of Saturn produces a torque on a satellite through a resonance lock,
\begin{eqnarray}
T_{\mathrm{mig}}&=&{1\over 2} m_s(GM_S a)^{1/2}\left(\dot a\over a\right)\,\nonumber\\
&&\nonumber\\
&&\times \left\{\begin{array}{ll}
1+{m_{\mathrm{out}}\over m_s}\left(a_{\mathrm{out}}\over a\right)^{1/2}\, ,&\mathrm{inner\,\, satellite\,\, in\,\, MMR\, ;} \\
0\, ,&\mathrm{outer\,\, satellite\,\, in\,\, MMR\, ;}\\
1\, ,& \mathrm{satellite\,\, not\,\, in\,\, MMR\, ,}
\end{array}\right.\nonumber\\
\label{eq:Tmig}
\end{eqnarray}
where $m_s$ and $a$ denote the mass and orbital semi-major axis of the satellite in a resonance lock, and $m_{\mathrm{out}}$ and $a_{\mathrm{out}}$ the mass and orbital semi-major axis of the outer satellite in the MMR. 

Although it is not involved in the resonance lock, the outer satellite in the MMR still raises a tidal bulge on Saturn. This is often called the equilibrium tide, although in a neutrally stratified body it is not equivalent to the conventionally defined equilibrium tide, as discussed in Section~\ref{sec:Phi-inertial-wave}. Small co-orbital satellites of Tethys and Dione are used by \citet{Lainey} to constrain the gravitational potential created by the tidal bulges induced by Tethys and Dione respectively. They are consistent with what is expected theoretically. Unfortunately, these two satellites are not in a resonance lock with Saturn, since they are both outer satellites in their respective MMRs.

\section{Gravitational potential perturbations}\label{sec:gravity-anomaly}

An oscillation of Saturn perturbs its external gravitational potential because the density field is perturbed. Even for the same torque provided by a resonance lock, a g mode and inertial wave attractor result in distinct gravitational potentials.

\subsection{Gravity modes}\label{sec:Phi-g-mode}

We assume stable stratification to reside between radii $r_c<r<\rb$, where $\rb$ is the bottom of the outer convective zone. \cite{Fuller-2014} propose a model with $r_c\approx 0.1R_S$ and $r_b\approx 0.4R_S$. The angular frequency of a g mode excited by a satellite in the rest frame of Saturn is $\sigma=m(\OS-\norb)$, where $m$ is the azimuthal order. Since $\sigma\sim \OS$, the Coriolis force strongly influences the angular pattern of a g mode by restricting horizontal motion\footnote{Vertical motion is predominantly controlled by gravity and pressure in stably stratified layers, and thus the Coriolis force is neglected in the radial direction.}. In the traditional approximation, Hough functions, rather than spherical harmonic functions, are the eigenfunctions of the angular part of the equations of motion \citep[e.g.][]{Chapman}\footnote{We constrain ourselves to Hough function of the first kind. Satellites orbit almost in the equatorial plane of Saturn, favoring excitation of modes concentrated toward the equator. Hough functions of the second kind concentrate toward the poles, as $|\sigma/(2\OS)|$ is close to unity \citep{Longuet-Higgins}, which is the case of interest in this paper. Modes with $n=m+1$, $m+3$, etc., are anti-symmetric about Saturn's equator, and therefore are not excited. Modes with $n=-(m+1)$, $-(m+2)$, $-(m+3)$, $-(m+4)$, etc, are Hough functions of the second kind.}. Hough functions are quantized by their angular degree, $n$\footnote{Hough functions' $n$ is analogous to $l$ for spherical harmonics.}, and their azimuthal order, $m$.  Modes excited by the tidal potential of a satellite must have their azimuthal order match that of the tidal potential field. The allowed values are $n=m,\,\, m+2, \,\, m+4$, etc. The corresponding g modes are closely packed in the  frequency domain. In other words, rotation makes the number of g modes per unit frequency larger than in the non-rotating case \citep[refer to Figure~1 in\,\,][]{Fuller-2016}. One important consequence is that satellites have more chances to encounter g modes in the frequency domain, which is a precondition favoring resonance locking.

Consider a g mode with $n_r$ radial nodes in the stably stratified region. The corresponding gravitational potential perturbation, $\Phi_g$, is dominated by the outermost half wavelength between $\rb-\lambda_1< r < \rb$, called the first half wavelength. Other half wavelengths partially cancel the potential perturbation generated by the first one but by at most $50\%$. The evanescent zone, i.e., the outer convective region, has no radial nodes in it. Its contribution to $\Phi_g$ has the same sign as that contributed by the first half wavelength, and thus strengthens the latter. These claims are justified in Appendix~\ref{sec:first-domination}. Below we estimate the potential perturbation due to the first half wavelength, $\Phifirst$. We simplify the first half wavelength by collapsing its density perturbation onto a layer at $\rb$. Poisson's equation reduces to
\begin{eqnarray}
\nabla^2\Phifirst &=& 4\pi G\rho^\prime(r,\theta,\varphi;t)\,\nonumber\\
&=& 4\pi G\Sigma_1\delta (r-\rb)\Theta_{nm}(\theta) e^{i (\sigma_m\, t+m\varphi)}\, ,
\end{eqnarray}
where $\rho^\prime$ is the Eulerian perturbation of density, $\Sigma_1$ is the column density perturbation, $\theta$ and $\varphi$ are the colatitude and longitude in the rest frame of Saturn, $\Theta_{nm}(\theta)$ is the Hough function with azimuthal order, $m$, and latitudinal degree $n$, $\delta(r)$ is the Dirac delta function, and 
\begin{equation}
\sigma_m\equiv m(\OS-\norb)\, .
\end{equation}
The eigenfunctions of the angular part of the Laplace operator in spherical coordinates are $\bar P_{lm}(\cos\theta)\exp(im\varphi)$, where $\bar P_{lm}(x)$ is a normalized associated Legendre polynomial. 
In order to solve for $\Phifirst$, we first project the Hough function, $\Theta_{nm}$, onto $\bar P_{lm}$. In this paper, we only consider the case $m=2$, because the leading order of the tidal field of the satellite is its quadrupole with $l=m=2$ \citep[e.g.][]{Murray-Dermott}. From now on we omit the index $m$, unless otherwise mentioned. We expand
\begin{equation}
\Theta_n(\theta) = \sum_{l\geq 2}^{\mathrm{even\,} l} \mathcal{B}_{nl} \bar P_l(\cos\theta)\, ,
\end{equation}
where we adopt the expansion coefficients, $\mathcal{B}_{nl}$, in Table (31) in \cite{Flattery}. Then the solution for $\Phifirst$ is also represented in the form of an expansion,
\begin{eqnarray}
\Phifirst(r> \rb)&=&\sum_{l\geq 2}^{\mathrm{even\,}l}\Phi_{nl}\bar P_l(\cos\theta)\exp(i\sigma t+i2\varphi)\, ,\label{eq:Phifirst}
\end{eqnarray}
where 
\begin{equation}\label{eq:Phinl}
\Phi_{nl}\equiv -4\pi G \Sigma_1 \rb {\mathcal{B}_{nl}\over (2l+1)} \left(\rb\over r\right)^{l+1}\, .
\end{equation}

A mode stores the same amount of energy, $E_{\mathrm{node}}$, between each pair of consecutive radial nodes. The total energy of a g mode with $n_r$ nodes is $(n_r+1)E_{\mathrm{node}}$, assuming that the evanescent region stores energy $\sim \Enode$. The mode energy gets damped at the rate, $\gamma (n_r+1)\Enode$. Correspondingly, the negative angular momentum carried by a retrograde mode gets damped at the rate, $\gamma (n_r+1)\Enode /\omega_p$, where $\omega_p=-\sigma/m=-\OS+\norb\approx -\OS$ is the azimuthal phase speed of the retrograde mode. In the equilibrium state of a resonance lock, i.e., not during capture or breaking of a resonance lock, the mode keeps its energy and angular momentum constant. Thus, the damped negative angular momentum of the mode must be replenished by the satellite. Because angular momentum is an invariant between rotating and inertial frames \citep{Pierce}, the satellite gains positive angular momentum at the rate,
\begin{equation}\label{eq:Tg}
T_g=-(n_r+1)\Enode{\gamma\over \omega_p} \approx (n_r+1)\Enode{\gamma\over\OS}\, .
\end{equation}
The requisite torque to push a satellite migrating at a rate $\dot a/a$ is equation~\refnew{eq:Tmig}.

Equations~\refnew{eq:Phifirst} and \refnew{eq:Phinl} express $\Phifirst$ in terms of $\Sigma_1$. Equations~\refnew{eq:Tg} and \refnew{eq:Tmig} relate $\Enode$ and $\dot a/a$. 
Next, we relate $\Enode$ and $\Sigma_1$, which will enable us to express $\Phifirst$ in terms of $\dot a/a$.  We constrain ourselves to the most general relations for g modes, making our result least dependent on the specific model of stable stratification in Saturn. There are two reasons for doing so. First, stable stratification is hypothetical; second, the biggest uncertainty in our result is $\gamma$ for g modes, and thus it is not worth spending much effort on the details of models for stable stratification. Our derivations are based on the following three assumptions:
\begin{enumerate}
\item The Eulerian density perturbation is
\begin{equation}\label{eq:rho-prime}
\rho^\prime \sim {d\rho\over dr}\xi_r {\sigma^2\rho\over k_h^2 p}\, ,
\end{equation}
where $d\rho/dr$ is the gradient of the background density profile, $k_h$ is the horizontal wave number, and $\rho$ and $p$ are the unperturbed background density and pressure. The derivation of $\rho^\prime$ is in appendix~\ref{sec:rho-prime}. 

\item The displacement of a g mode is dominated by its horizontal component,
\begin{equation}
\xi_h\gg \xi_r\, .
\end{equation}

\item The bottom of the outer convective zone is the upper edge of the propagation cavity of a g mode. At this outer turning point, the wavelength of a g mode is comparable to the local scale height, where the WKB dispersion relation starts to break down. For simplicity, we do not distinguish a pressure scale height and a density scale height. Consequently, at $\rb$,
\begin{equation}\label{eq:kr}
k_r\sim {1\over\lambda_1} \sim {1\over H}\, .
\end{equation}
\end{enumerate}

The energy in each node of a g mode is then
\begin{equation}\label{eq:Enode}
\Enode\sim \rb^2\lambda_1 \rho_b \sigma^2 \xi_h^2 \sim \rb^2\rho_b\sigma^2  {\xi_r^2 k_r \over  k_h^2}\, .
\end{equation}
The column density perturbation of the first half wavelength is
\begin{equation}\label{eq:Sigma1-1}
\Sigma_1\sim\rho^\prime \lambda_1\, , 
\end{equation}
where we have adopted $d\rho/dr \sim \rho_b/H$. Eliminating $\xi_r$ in equations~\refnew{eq:Enode} and \refnew{eq:Sigma1-1}, we obtain 
\begin{eqnarray}
\Sigma_1 &\sim &\, {r_b^2 \lambda_1 \xi_r \rho_b \sigma^2 \over g_b H^2 K_n}\label{eq:Sigma1-2}\, \\
\nonumber\\
&\sim & {\Enode^{1/2} \rho_b^{1/2} \sigma\over g_b H^{1/2} n}\, ,\label{eq:Sigma1-3}
\end{eqnarray}
where $g_b$ is the gravity at $r_b$, $\bar\rho\sim M_S/R_S^3$, and we have expressed the horizontal wavenumber\, ,
\begin{equation}
k_h = {K_n^{1/2}\over r} \approx {n\over r}\, .
\end{equation}
This approximation holds true if $4\OS^2 /(g k_r) \ll 1$ \citep{Longuet-Higgins}. For azimuthal order, $m=2$, $n$ has to be even and $\geq2$, since we only consider Hough functions of the first kind. Combining equations~\refnew{eq:Tmig}, \refnew{eq:Phinl}, \refnew{eq:Tg} and  \refnew{eq:Sigma1-3} and eliminating $\Enode$ and $\Sigma_1$, we obtain the gravity anomaly,
\begin{eqnarray}
{\Phi_{nl}\over (-GM_S/R_S)}&\sim & {\mathcal{B}_{nl}\over (2l+1)}\left(R_S\over a\right)^2\left(r_b\over R_S\right)^{l+2}\left(R_S\over r\right)^{l+1}\left(g_S\over g_b\right)\nonumber\\
&&\times \left(\Omega_S\over\Omega_{\mathrm{orb}}\right)^{3/2}\left(\dot a/a\over (n_r+1)\gamma\right)^{1/2}\left(m_s\over M_S\right)^{1/2}\,\nonumber\\
&& \times \left(\rho_b\over \bar\rho\right)^{1/2}\left(R_S \over K_n H\right)^{1/2}\, ,\nonumber\\
\label{eq:Phinl-final}
\end{eqnarray}
where $g_S\equiv GM_S/R_S^2$. We have replaced $\sigma$ by $2\Omega_S$ for azimuthal order of $2$ which is the case we consider here.
Note that the above expression corresponds to a satellite not in a MMR, e.g., Rhea. The outer satellite in a MMR, such as Tethys and Dione, is not involved in a resonance lock, and thus there is no $\Phi_{nl}$. For the inner satellite in a MMR, namely Mimas, Enceladus and Titan, we need to replace $\dot a/a$ by the following factor,
\begin{equation}\label{eq:replacement}
\left(\dot a\over a\right)\left(1+{m_{\mathrm{out}}\over m_s}\left(a_{\mathrm{out}}\over a\right)^{1/2}\right)\, .
\end{equation}

The analysis above assumes linear damping. However, at large $k_r\xi_r$, nonlinear damping sets in and limits the amplitude. Different nonlinear effects, e.g. three-mode coupling \citep[e.g.][]{Wu, Weinberg}, generation of turbulence \citep[e.g.][]{Hodges}, wave front steepening \citep[e.g.][]{Greenspan-1958}, etc., could limit $k_r \xi_r$ to different thresholds. However, we do not know which nonlinear effect dominates, without good knowledge of the stratification in Saturn. Without any better criterion, we employ 
\begin{equation}
\left. k_r\xi_r\right|_{r_b}\sim 1
\end{equation}
as the threshold over which nonlinear damping would limit the amplitude. Combining equations~\refnew{eq:Tmig}, \refnew{eq:Tg}, \refnew{eq:kr} and \refnew{eq:Enode}, we obtain
\begin{eqnarray}
\left. k_r\xi_r\right|_{r_b}&\sim & n \left(\dot a/a \over (n_r+1)\gamma\right)^{1/2}\left(m_s\over M_S\right)^{1/2}\left(\Omega_{\mathrm{orb}}\over \Omega_S\right)^{1/2} \,\nonumber\\
&&\times \left(R_S\over H\right)^{1/2}\left(R_S\over r_b\right)^2\left(a\over R_S\right)\left(\bar\rho\over \rho_b\right)^{1/2} \, .\label{eq:krxir}
\end{eqnarray}
Similarly, for inner satellites in a MMR, $\dot a/a$ should be replaced by equation~\refnew{eq:replacement}.

\subsection{Inertial wave attractors}\label{sec:Phi-inertial-wave}

This subsection mainly quotes Section 4 in \cite{Ogilvie-2013}\footnote{We review Ogilvie's result with our own understanding. Readers interested in details are referred to \cite{Ogilvie-2013}. For general readers, it suffices to read this subsection.}, which studies slow oscillations in slowly rotating barotropic bodies with $\sigma<2\OS\ll (GM_S/R_S^3)^{1/2}$. A barotropic fluid has its pressure uniquely related to density and therefore is neutrally stratified. These conditions apply for inertial waves in the outer convective region of Saturn. The wave displacement is decomposed into a non-wave-like part, $\bmath{\xi}_{\mathrm{nw}}$, and a wave-like part, $\bmath{\xi}_{\mathrm{w}}$. The former is the instantaneous hydrostatic response of the fluid to the external tidal force from the satellite, while the latter is driven by the unbalanced Coriolis force induced by the former, $-2\bmath{\OS}\times \dot{\bmath{\xi}}_{\mathrm{nw}}$. Note that the instantaneous hydrostatic response, $\bmath{\xi}_{\mathrm{nw}}$, is conceptually similar but not equivalent to the conventional equilibrium tide. The equilibrium tide represents the tidal response in the zero frequency limit and is not well defined in the absence of stable stratification. Since Saturn is believed to be mostly convective, the non-wave-like hydrostatic response instead of the equilibrium tide is the proper term here, although sometimes people do not distinguish them. 

The non-wave like part, $\bmath{\xi}_{\mathrm{nw}}$, generates a gravitational potential proportional to that of the satellite,
\begin{eqnarray}
\Phi_{\mathrm{nw}}(r>R_S)& =& \sum_{l,m} k_{lm} \Psi_{lm}\left(R_S\over r\right)^{l+1}\nonumber\\
&&\,\,\,\,\,\,\,\,\, \times \bar P_{lm}(\cos\theta_i)\exp(im(\varphi_i-\norb t))\, ,\nonumber\\
\label{eq:Phinw}
\end{eqnarray}
using the expansion of the tidal potential of the satellite
\begin{eqnarray}
\Phi_{\mathrm{ext}} &=&  \sum_{l,m} \Psi_{lm}\left(r\over a\right)^{l} \bar P_{lm}(\cos\theta_i)\nonumber\\
&&\,\,\,\,\,\,\,\,\, \times \exp(im(\varphi_i-\norb t))\, ,
\end{eqnarray}
where the sum is over integers $l\geq 2$ and $-l\leq m\leq l$, $\theta_i$ and $\varphi_i$ are the colatitude and longitude in the inertial frame , and
\begin{equation}
\Psi_{lm}\approx - {G m_s \over a}\left(R_S\over a\right)^l\, .
\end{equation}
The pattern speed in the inertial frame, i.e., the azimuthal phase speed, is $\norb$. The Love number, $k_{lm}$, is usually complex. Dissipative processes lead to ${\it Im}(k_{lm})$ that is usually small compared to ${\it Re}(k_{lm})$. The imaginary part causes a small phase lag between $\Phi_{\mathrm{nw}}$ and $\Phi_{\mathrm{ext}}$. 

The tidal Love numbers, $k_{lm}$, depend on the internal structure of Saturn, especially its density distribution. Density profiles increasing toward the center usually produce small Love numbers because the tidal force vanishes at the center of planet. Love numbers also depend on the frequency in the rest frame of Saturn, $m(\OS-\norb)$, \citep[e.g.][]{Goodman-Lackner, Ogilvie-2013}. Note that the conventionally defined equilibrium tide, \citep[e.g.][]{Goldreich-Nicholson-1989a, Goldreich-Nicholson-1989b}, is frequency independent because it is calculated assuming $\partial/\partial t=0$ so that any time or frequency dependence is erased. 

The wave-like part, $\bmath{\xi}_{\mathrm{w}}$, is an inertial wave. It does not generate a gravitational potential perturbation in the limit $\sigma<2\OS\ll (GM_S/R_S^3)^{1/2}$, because inertial waves lack the ability to raise a free surface \citep{Ogilvie-2013}. According to equations (44) and (46) and discussion below them in \cite{Ogilvie-2013}, the potential corresponding to the wave-like part is on the order of $\OS^2R_S^3/(GM_S)\Psi_{\mathrm{nw}}\sim 0.1\Psi_{\mathrm{nw}}$, i.e. about 10 times smaller than that contributed by the non-wave like part. Therefore, if it were an inertial wave attractor that resonantly locks a satellite, the corresponding gravitational perturbation would be nearly the same as that generated by the non-wave like part of the tidal response. Such a small difference would be difficult to distinguish from uncertainties in $\Psi_{\mathrm{nw}}$ due to uncertainties in Saturn's interior structure.

\section{Comparison with expected accuracy of gravity measurement by {\it Cassini}}\label{sec:compare}

We mentioned at the end of the introduction that {\it Cassini} will measure the gravitational field of Saturn as it flies by its surface. The anticipated one sigma accuracy for the gravity coefficients, $J_l$, after 6 Proximal orbits, through private communication with Phillip D. Nicholson, are
\begin{eqnarray}
\Delta_2 &=& 2\times 10^{-9}\, ,\\
\Delta_6 &=& 2\times 10^{-8}\, ,\\
\Delta_{10} &=& 1\times 10^{-7}\, ,\\
\Delta_{14} &=& 2\times 10^{-7}\, ,
\end{eqnarray}
given that the gravitational potential of Saturn is expanded as
\begin{equation}
\Phi_S (r>R_S)= -{GM_S\over r}\left [ 1+\sum_{l=2}^{N_z}\left(R_S\over r\right)^l J_l P_{l0}(\cos\theta)\right ]\, , \label{eq:PhiS}
\end{equation}
where $r$ is the distance from the center of Saturn, $P_{lm}$ are the associated Legendre polynomials\footnote{Not normalized yet.}, and we do not include terms with $m\neq 0$ for brevity. At the time of writing this manuscript, Luciano Iess comments that the accuracy achieved after 4 Proximal orbits depends on the dynamic models for fitting the data, and is about 10 times worse than those quoted here. We still quote the optimal anticipated accuracies since {\it Cassini} is still collecting more data, and hopefully the accuracy could be improved. We speculate the accuracy declines at high orders $J_l$ because {\it Cassini} makes gravity measurement mainly at $r\sim 2R_S$\footnote{Private communication with Phillip D. Nicholson.} and the gravitational potential corresponding to higher orders decays faster with distance. Both the potential component associated with $J_l$ and that due to a g mode, $\Phi_{nl}$, declines as $r^{-(l+1)}$. We regard $\Delta_l$ as the accuracy of the coefficient associated with the component of the gravitational potential which decays with the $l+1$ power of distance.

\begin{figure}
\includegraphics[width=0.9\linewidth]{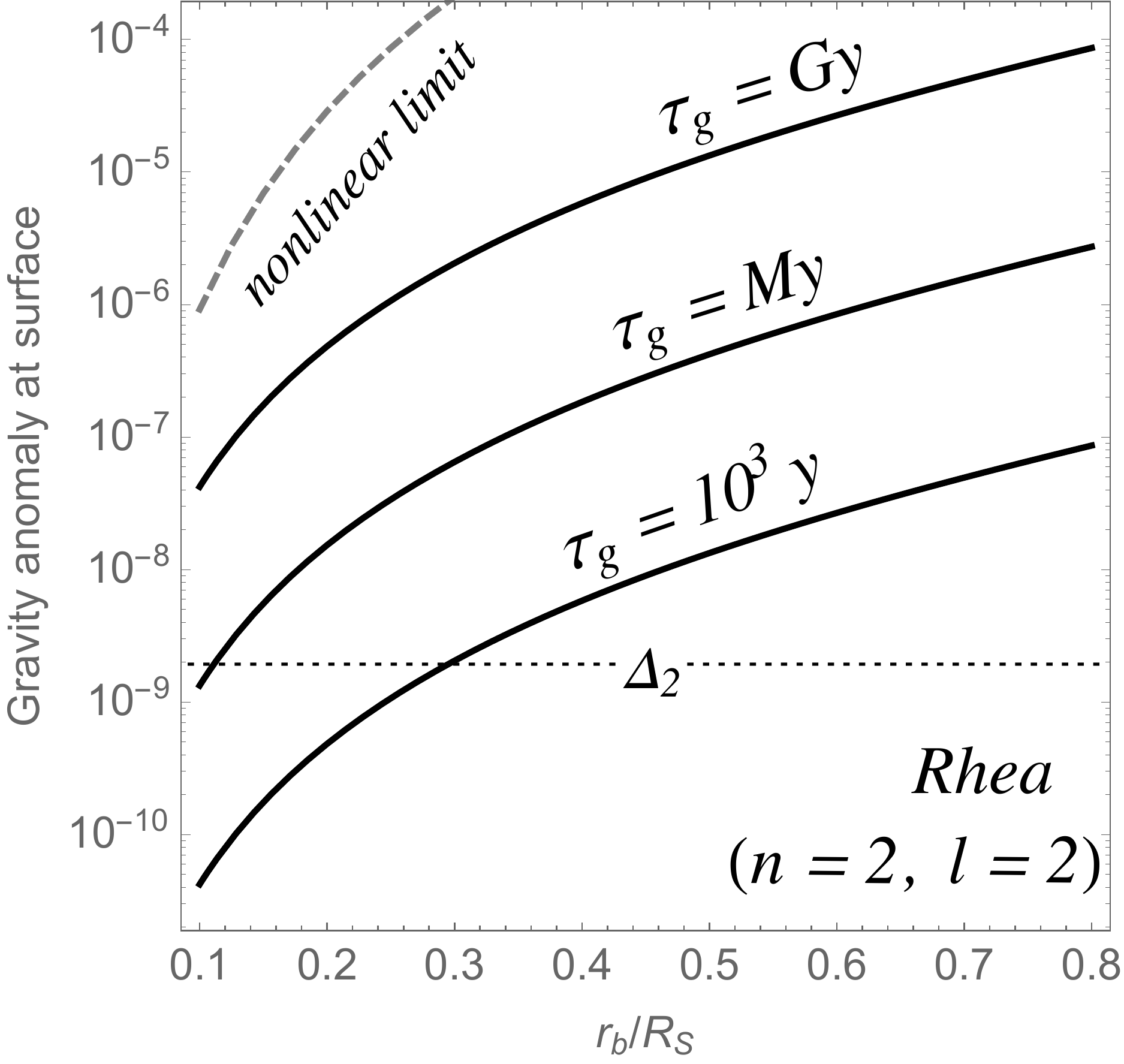}
\caption{\label{fig:Rhea-22-surf} Gravity anomaly at the surface of Saturn, equation~\refnew{eq:anomaly-nl}, for a g mode with $l=2$ and $n=2$ that is resonantly locked with Rhea. The horizontal axis is $\rb/R_S$, where $\rb$ is the bottom of the outer convection region and the top of the hypothetical stably stratified region. The corresponding mode damping time, $\tau_g$, is labeled beside each curve. The dotted horizontal line shows the measurement sensitivity of {\it Cassini}. The dashed curve corresponds to $(k_r\xi_r)_{r_b}\sim 1$. Nonlinear damping limits the gravity anomaly to lie below this line.  }
\end{figure}

Comparing the format of $\Phifirst$ in equation~\refnew{eq:Phifirst} and that of $\Phi_S$ in equation~\refnew{eq:PhiS}, we realize that $\Phi_{nl}(r=R_S)$ is associated with $(R_S/r)^l \bar P_{l,m=2}(\cos\theta)$, whereas $(-GM_S/R_S) J_{l}$ is associated with $(R_S/r)^lP_{l,m=0}(\cos\theta)$. We speculate that the accuracy for $\Phi_{nl}(r=R_S)/(-GM_S/R_S)$ is similar to $\Delta_l$. In other words, we speculate the non-axisymmetric component of Saturn's gravitational field due to Saturn's tidal deformations can be measured with similar precision to its axisymmetric gravitational field. We calculate
\begin{eqnarray}
{\Phi_{nl}(r=R_S)\over (-GM_S/R_S)}&\sim& {\mathcal{B}_{nl}\over n (2l+1)}\left(r_b\over R_S\right)^{l+2}\left(R_S\over a\right)^2\,\nonumber\\
&& \times \left(m_s\over M_S\right)^{1/2}\left(\rho_b\over \bar\rho\right)^{1/2}\left(\dot a/a\over (n_r+1)\gamma\right)^{1/2}\,\nonumber\\
&& \left(\Omega_S\over\Omega_{\mathrm{orb}}\right)^{3/2} \times \left(g_S\over g_b\right)\left(R_S\over H\right)^{1/2}\, ,\label{eq:anomaly-nl}
\end{eqnarray}
Note that $\dot a/a$ needs to be replaced by equation~\refnew{eq:replacement} for the inner satellite in a MMR. 
We show an example of the gravity anomaly at the surface of Saturn, $\Phi_{22}(R_S)/(-GM_S/R_S)$, generated by a g mode resonantly locked with Rhea in figure~\refnew{fig:Rhea-22-surf}. It is above the measurement sensitivity of {\it Cassini} for most of the range of $r_b/R_S$. 

Linear damping due to turbulent viscosity or heat diffusion yields $\gamma^{-1}\sim (1\sim 10)\,\mathrm{Gyr}$. A gravity mode with azimuthal order $m=2$ has its angular frequency $\sigma=2(\OS-\norb)$, within the range for inertial waves, $(-2\OS, 2\OS)$. Therefore, g modes may suffer additional damping upon conversion to inertial waves in the convection zone \citep[e.g.][]{Dintrans,Mathis}. Nevertheless, the frequencies of g modes do not generally coincide with frequencies of inertial wave attractors. Considering that inertial waves do not suffer significant damping unless forming an attractor \citep{Ogilvie-Lin, Ogilvie-2013}, we speculate that $\gamma$ does not greatly increase through g mode coupling with inertial waves in the convection zone. There may also exist other damping mechanisms beyond our knowledge. We must acknowledge that $\gamma$ is very uncertain, and we decide to leave it as a free parameter. Fortunately, $\Phi_{nl}\propto \gamma^{-1/2}$, depending only weakly on $\gamma$. The square root dependence can be understood in the following way. The tidal torque on a satellite scales as $T_{\mathrm{osc}}\propto \gamma A^2$, while the gravitational potential scales as $\Phi_{nl}\propto A\propto (T/\gamma)^{1/2}$. Since $T\propto \dot a/a$, we also have $\Phi_{nl}\propto (\dot a/a)^{1/2}$.

On the other hand, $\Phi_{nl}$ depends on $r_b/R_S$ most sensitively.  \cite{Fuller-2014} propose that $r_b/R_S\approx 0.4$ but this value is uncertain. Besides the explicit dependence, $(r_b/R_S)^{(l+2)}$, $\rho_b$, $H$ and $g_b$ also depend on $r_b$. For a polytrope with index $1$ (Appendix~\ref{sec:gamma-turb}), we have $(\rho_b /H/g_b)^{1/2}$ approaches a constant for $r_b/R_S\ll 1$. Therefore,
\begin{equation}
\Phi_{nl}\appropto\left(r_b\over R_S\right)^{l+2}\, ,\, \, (r_b\ll R_S)\, ,
\end{equation}  
yielding that constant $\Phi_{nl}$ roughly traces 
\begin{equation}
\tau_g\equiv {1\over (n_r+1)\gamma}\propto  (r_b/R_S)^{-2l-4}\, ,\,\, (\mathrm{for}\,\, r_b\ll R_S)\, .\label{eq:taug}
\end{equation}
The sensitive dependence of $\Phi_{nl}$ on $(r_b/R_S)$ can be understood as follows. The power index $l$ originates from the decaying potential of the $l$th multipole with distance. The rest of the power index finds its root in the fact that the background density gradient flattens towards small $r_b/R_S$ for a polytrope with index unity. We show right below equation~\refnew{eq:rho-prime} that a flatter density gradient makes it harder for a gravity mode to perturb the gravitational potential. 

\begin{figure}
\includegraphics[width=0.8\columnwidth]{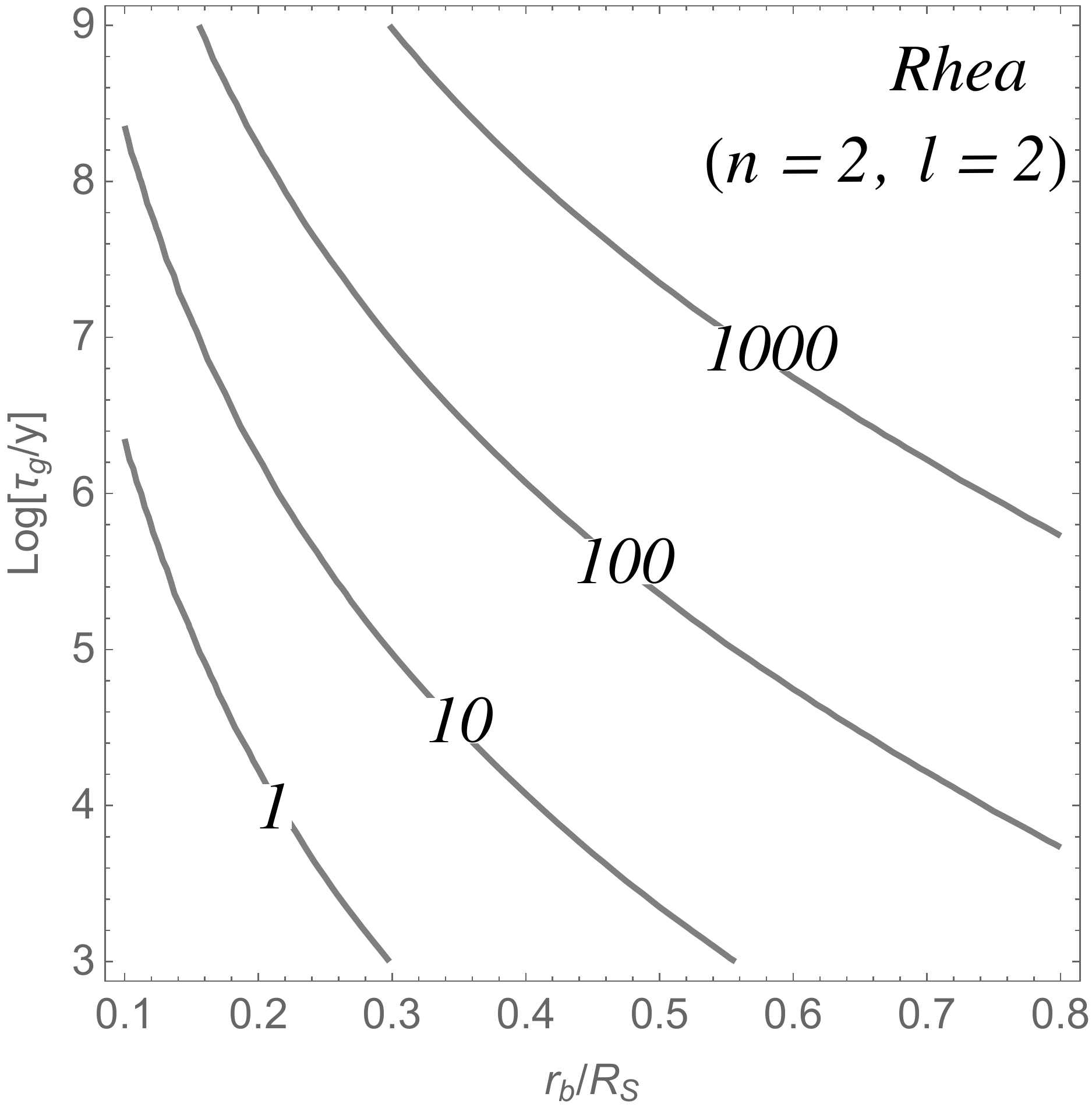}
\includegraphics[width=0.8\columnwidth]{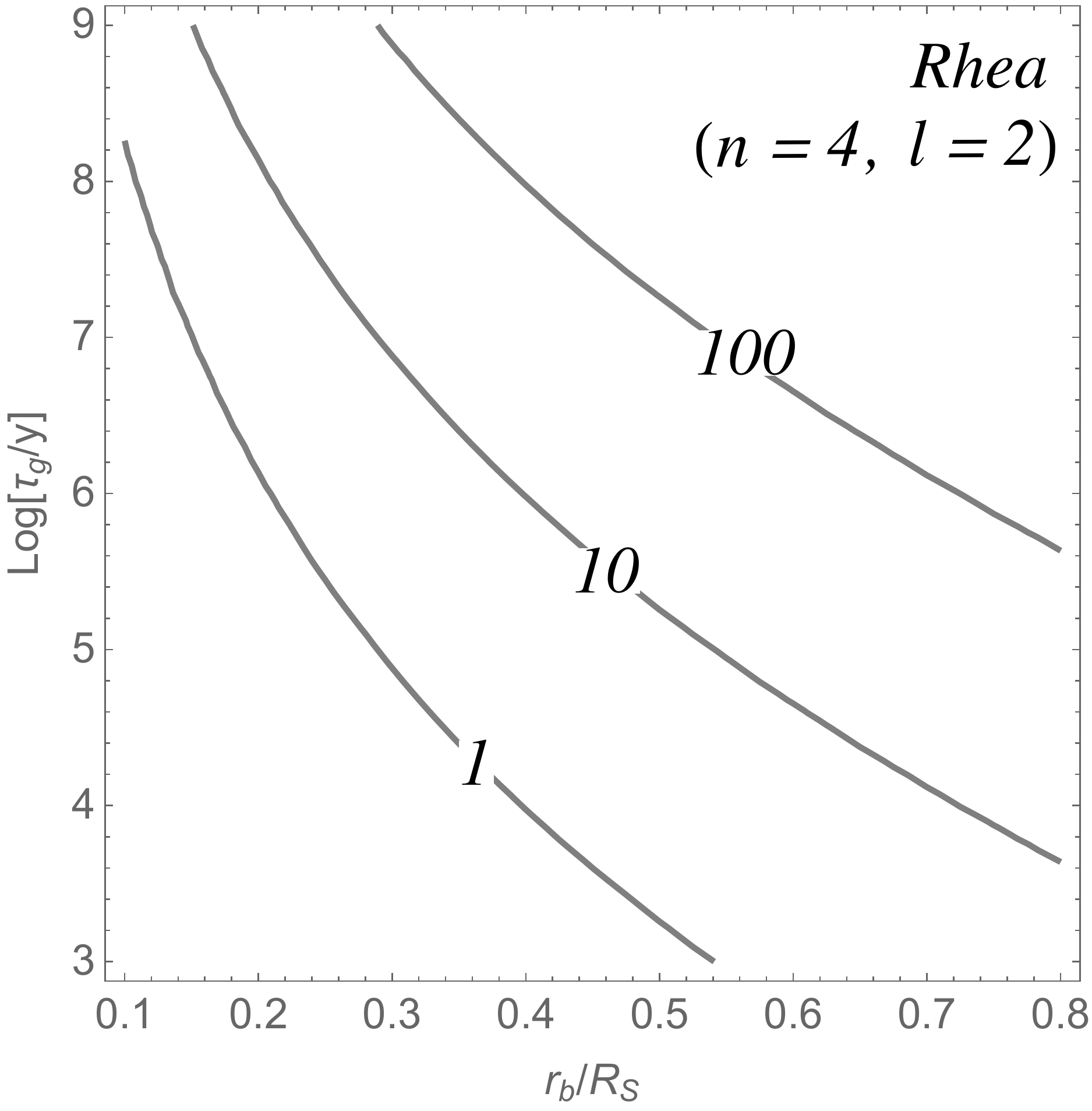}
\includegraphics[width=0.8\columnwidth]{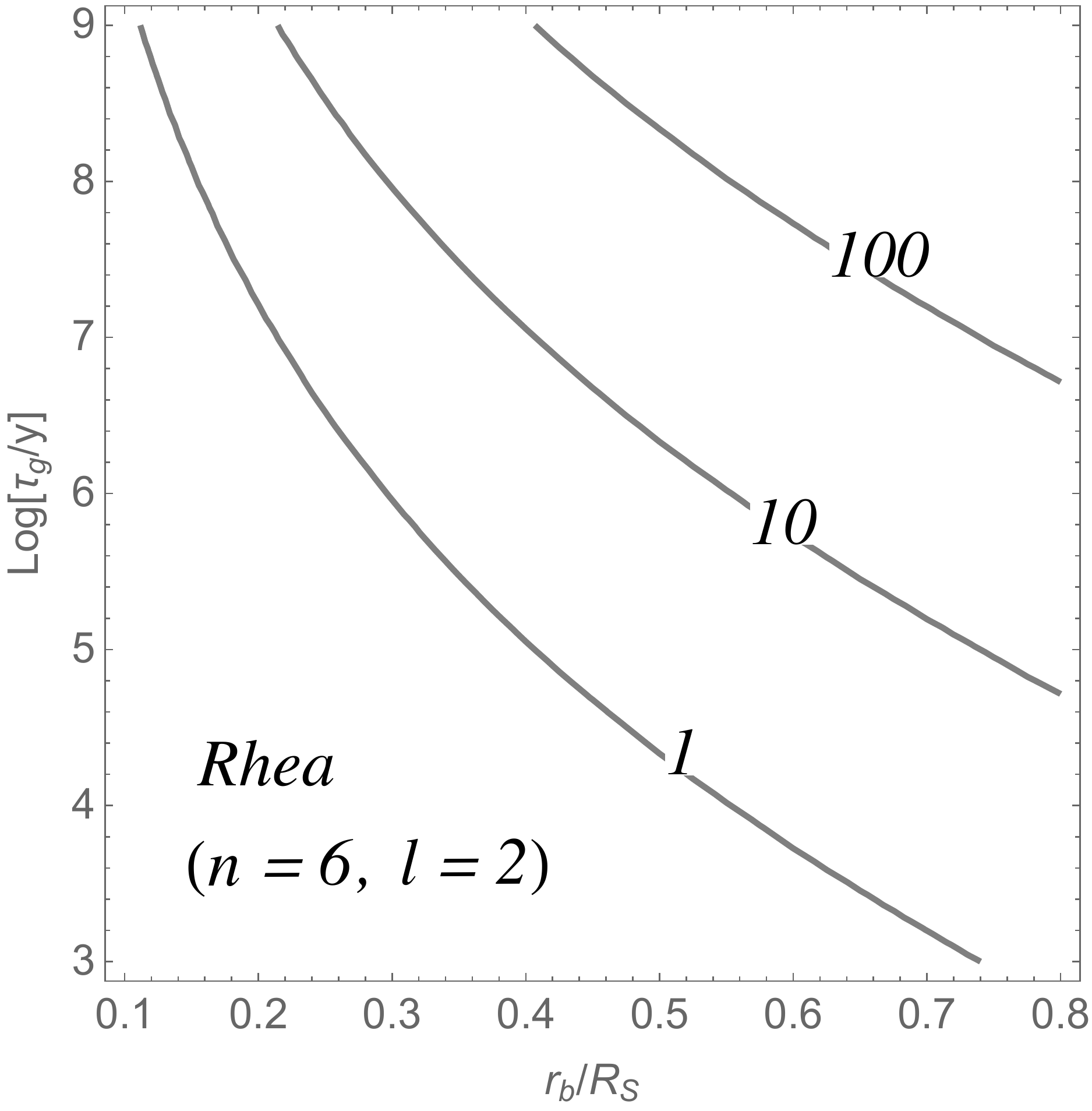}
\caption{\label{fig:Rhea}Parameter space in which we compare the gravitational perturbation of a g mode resonantly locked with Rhea with the sensitivity of {\it Cassini}. The vertical axis is the g mode damping time, $\tau_g\equiv 1/\gamma/(n_r+1)$, in units of years. The solid contour line labeled by `1' shows where the gravity anomaly produced by the resonant g mode, equation~\refnew{eq:anomaly-nl}, is equal to the expected sensitivity of {\it Cassini}'s gravity measurement, $\Delta_l$. Other solid contour lines show where the gravity anomaly is equal to $10$, $100$, etc. times $\Delta_l$. The nonlinear limit, $(k_r\xi_r)_{r_b}\sim 1$, lies above the parameter space shown here. }
\end{figure}

We illustrate a solid contour line with gravity anomaly (equation~\refnew{eq:anomaly-nl}) equal to $\Delta_l$ in figure~\refnew{fig:Rhea}, for Rhea with $l=2$ and $n=2,\,\, 4,\,\, 6$. 
The region above each contour line labeled `1' is where the gravity anomaly is greater than $\Delta_l$, where {\it Cassini} could hopefully detect the gravity potential from a resonantly locked g mode.  The limit due to nonlinear effects is above the parameter space shown in this figure. 

Gravitational perturbations for other satellites, e.g., Mimas and Enceladus, are similar to figure~\refnew{fig:Rhea}. Note that we only consider $l=2$ because the measurement accuracy is best at $l=2$. Assuming Titan also migrates at a rate $\dot a/a\sim 1/(5\,\mathrm{Gy})$, it would create a gravity anomaly $\sim \! 10$ times larger (at the same $r_b/R_S$ and $\tau_g$) than Rhea.

If inertial wave attractors resonantly lock satellites, then the gravitational potential perturbation is generated by the non-wave-like response of Saturn. The format of $\Phi_S$ in equation~\refnew{eq:PhiS} and $\Phi_{nw}$ in equation~\refnew{eq:Phinw} suggest that we compare $\Delta_l$ with
\begin{equation}
{k_{lm}\Psi_{lm}\over (-GM_S/R_S)}\approx k_{lm}\left(m_s\over M_S\right)\left(R_S\over a\right)^{l+1}\, .
\end{equation}
This evaluates to $\approx 2.6\times 10^{-8} k_{lm}$ for Titan with $l=2$, which may be detectable. For higher order multipoles and other other satellites, the gravitational perturbations from inertial waves are likely too weak to detect. 

The gravitational potential of Saturn is also affected by other types of perturbations, e.g., zonal winds, which may dominate the $J_l$'s. However, the gravitational potential generated by the tidal response of Saturn has its azimuthal phase speed equal to the mean motion of the satellite. Therefore, a frequency analysis can filter out other types of perturbations because they will not follow the same pattern speed as the satellite.

\section{Fundamental modes of Saturn}\label{sec:f-mode-main-body}

\cite{Hedman-Nicholson} report density waves in the C ring of Saturn that propagate inward. They are believed to be generated by outer Lindblad resonances with fundamental modes (f modes) of Saturn. It is interesting to check whether the f modes are detectable through measuring their gravity anomaly. 
We estimate the gravitational potential of f modes in an order of magnitude way in appendix~\ref{sec:f-mode}. Additional numerical calculations can be found in \citet{Marley} and \citet{Fuller-2014}. The $l=m$ component of the perturbed gravitational potential is found to be
\begin{equation}
{\Delta\Phi_f \over (GM_S/R_S)} \sim {2\pi \over m(3m+1)} \left(r_L\over R_S\right)^m \left(\Sigma r_L^2\over M_S\right)\, .
\end{equation}
where $r_L$ is the location of the outer Lindblad resonance and $\Sigma$ is the mass column density at $r_L$. Estimates of both $r_L$ and $\Sigma$ are available in tables 4 and 6 in \cite{Hedman-Nicholson}. We compare the gravity anomaly due to a hypothetical f mode with $\Delta_m$ in table~\ref{tab:f-mode}. Unfortunately, the magnitudes of the gravity anomalies generated by f modes are below the anticipated measurement accuracy of {\it Cassini}. 

\begin{table}
	\centering
	\caption{Potential perturbation due to f modes at $R_S$, which is compared with the accuracy of {\it Cassini}'s gravity measurement, $\Delta_m$. We adopt the same notation for the names of each f mode as those in \citet{Hedman-Nicholson}. Here $\Delta_m$ is the sensitivity for the coefficient associated with the potential component decaying by the $m+1$ power of distance. Given that $\Delta_2\sim 2\times 10^{-9}$ and $\Delta_6\sim 2\times 10^{-8}$ (private communication with Philipp D. Nicholson), we assume that $\Delta_3\sim \Delta_4\sim \sqrt{\Delta_2\Delta_6}\sim 6\times 10^{-8}$. }
	\label{tab:f-mode}
	\begin{tabular}{lcccr} 
		\hline
		\hline
		Wave & m & $r_L\,(\mathrm{km})$ & $\Delta \Phi_f/(GM_S/R_S)$ & {\it Cassini} accuracy\\
		\hline
		W80.98 & 4 & 80988 & $3.1\times 10^{-10}$ & $ 6\times 10^{-8}$ \\
		W82.00 & 3 & 82010 & $4.0\times 10^{-10}$  & $ 6\times 10^{-8}$ \\
		W82.06 & 3 & 82061 & $7.1\times 10^{-10}$ & $6\times 10^{-8}$ \\
		W82.21 & 3 & 82209 & $4.9\times 10^{-10}$ & $ 6\times 10^{-8}$\\
		W84.64 & 2 & 84644 & $4.8\times 10^{-10}$ & $ 2\times 10^{-9}$ \\
		W87.19 & 2 & 87189 & $1.9\times 10^{-10}$ & $ 2\times 10^{-9}$ \\
		\hline
	\end{tabular}
\end{table}

The ring system of Saturn is a sensitive seismometer. The discussion above about f modes suggests that Saturn's C ring is an even more sensitive seismometer than {\it Cassini}. Oscillations of Saturn and gravitational forcing from satellites both excite density waves or bending waves in the ring, as long as their azimuthal phase speed in the inertial frame matches an integer ratio, $p/q$, multiplied by the local Keplerian orbital frequency in the ring \citep[e.g.][]{Hedman-Nicholson, Nicholson-I, Nicholson-II, French}. However, oscillations of Saturn excited by satellites share the same azimuthal phase speed as the satellite, therefore they would produce ring waves at the same location as waves excited directly by the satellite. Unless there is good knowledge about the properties of the rings \citep[e.g.][]{Hedman, Spilker} that precisely constrains the strength of perturbing gravitational potential, the ring system may not distinguish between a tidally excited oscillation of Saturn and the satellite itself.

\section{Conclusion}\label{sec:conclusion}

Resonance locking between satellites and oscillations of Saturn is a promising mechanism to account for the current surprisingly fast migration of satellites \citep{Lainey}. We investigate two channels through which resonance locking can operate. One is to increase the amplitude of an oscillation while keeping the damping rate constant, e.g., a g mode; the other is to increase the damping rate while keeping the amplitude constant, e.g., an inertial wave attractor. 

Since the perturbation to the gravitational field of Saturn by an oscillation is proportional to its amplitude, g modes and inertial wave attractors would generate distinctive gravity potentials. The {\it Cassini} spacecraft will finish its proximal orbits by September 2017. Flying by Saturn closely multiple times, it will constrain the gravity field of Saturn to unprecedented accuracy. According to our estimates, the gravitational potential produced by a g mode resonantly locked with a satellite is detectable for a wide range of the parameter space. Because the azimuthal phase speed of a g mode resonantly locked with a satellite is well known, filtering out the gravity anomaly at the corresponding frequency would largely damp noise due to other reasons, e.g., zonal winds. Additionally, the gravitational perturbation due to a resonantly locked g mode is very sensitive to the depth of the convective region in Saturn. In spite of the uncertain in damping rates of g modes, detection of their gravitational perturbations would provide constraints on the depth of the convective region. 

On the other hand, the gravitational perturbations caused by resonantly locked inertial wave attractors are likely too small to detect. However, given the fact that g modes are detectable in a large region of parameter space, a null detection of g modes would favor the operation of inertial wave attractors. The frequencies at which attractors form depend sensitively on the size of the cavity inside which inertial waves are confined. So, the frequencies of inertial wave attractors can also put meaningful constraints on the interior structure of Saturn.

\section*{Acknowledgements}
We thank Peter Goldreich for his insightful suggestions and comments. We thank Phillip D. Nicholson for providing us information about the proximal orbits of {\it Cassini}. We thank Gordon Ogilvie for his clarification about his work on inertial waves, i.e. \cite{Ogilvie-2013}. We thank Douglas N C Lin for his comments on inertial wave attractors. This work is supported by the Theoretical Astronomy Center and Center for Integrative Planetary Science at University of California at Berkeley. This research is funded in part by the Gordon and Betty Moore Foundation through Grant GBMF5076 and by the Simons Foundation through a Simons Investigator Award to EQ.




\bibliographystyle{mnras}
\bibliography{Gravity-anomaly-mnras} 

\begin{thebibliography}{}
\makeatletter
\relax
\def\mn@urlcharsother{\let\do\@makeother \do\$\do\&\do\#\do\^\do\_\do\%\do\~}
\def\mn@doi{\begingroup\mn@urlcharsother \@ifnextchar [ {\mn@doi@}
  {\mn@doi@[]}}
\def\mn@doi@[#1]#2{\def\@tempa{#1}\ifx\@tempa\@empty \href
  {http://dx.doi.org/#2} {doi:#2}\else \href {http://dx.doi.org/#2} {#1}\fi
  \endgroup}
\def\mn@eprint#1#2{\mn@eprint@#1:#2::\@nil}
\def\mn@eprint@arXiv#1{\href {http://arxiv.org/abs/#1} {{\tt arXiv:#1}}}
\def\mn@eprint@dblp#1{\href {http://dblp.uni-trier.de/rec/bibtex/#1.xml}
  {dblp:#1}}
\def\mn@eprint@#1:#2:#3:#4\@nil{\def\@tempa {#1}\def\@tempb {#2}\def\@tempc
  {#3}\ifx \@tempc \@empty \let \@tempc \@tempb \let \@tempb \@tempa \fi \ifx
  \@tempb \@empty \def\@tempb {arXiv}\fi \@ifundefined
  {mn@eprint@\@tempb}{\@tempb:\@tempc}{\expandafter \expandafter \csname
  mn@eprint@\@tempb\endcsname \expandafter{\@tempc}}}

\bibitem[\protect\citeauthoryear{{Barker}, {Dempsey}  \& {Lithwick}}{{Barker}
  et~al.}{2014}]{Barker}
{Barker} A.~J.,  {Dempsey} A.~M.,   {Lithwick} Y.,  2014, \mn@doi [\apj]
  {10.1088/0004-637X/791/1/13}, \href
  {http://adsabs.harvard.edu/abs/2014ApJ...791...13B} {791, 13}

\bibitem[\protect\citeauthoryear{{Borderies}, {Goldreich}  \&
  {Tremaine}}{{Borderies} et~al.}{1982}]{Borderies}
{Borderies} N.,  {Goldreich} P.,   {Tremaine} S.,  1982, \mn@doi [\nat]
  {10.1038/299209a0}, \href {http://adsabs.harvard.edu/abs/1982Natur.299..209B}
  {299, 209}

\bibitem[\protect\citeauthoryear{{C{\'e}bron}, {Bars}, {Gal}, {Moutou},
  {Leconte}  \& {Sauret}}{{C{\'e}bron} et~al.}{2013}]{Cebron}
{C{\'e}bron} D.,  {Bars} M.~L.,  {Gal} P.~L.,  {Moutou} C.,  {Leconte} J.,
  {Sauret} A.,  2013, \mn@doi [\icarus] {10.1016/j.icarus.2012.12.017}, \href
  {http://adsabs.harvard.edu/abs/2013Icar..226.1642C} {226, 1642}

\bibitem[\protect\citeauthoryear{{Chapman} \& {Lindzen}}{{Chapman} \&
  {Lindzen}}{1970}]{Chapman}
{Chapman} S.,  {Lindzen} R.,  1970, {Atmospheric tides. Thermal and
  gravitational}.
D. Reidel Publishing Company, Dordrecht-Holland

\bibitem[\protect\citeauthoryear{{Charnoz} et~al.,}{{Charnoz}
  et~al.}{2011}]{Charnoz}
{Charnoz} S.,  et~al., 2011, \mn@doi [\icarus] {10.1016/j.icarus.2011.09.017},
  \href {http://adsabs.harvard.edu/abs/2011Icar..216..535C} {216, 535}

\bibitem[\protect\citeauthoryear{Cox}{Cox}{1980}]{Cox}
Cox J.,  1980, The Theory of Stellar Pulsation.
Princeton series in astrophysics, Princeton University Press, \url
  {https://books.google.com/books?id=-kiqQgAACAAJ}

\bibitem[\protect\citeauthoryear{{{\'C}uk}, {Dones}  \&
  {Nesvorn{\'y}}}{{{\'C}uk} et~al.}{2016}]{Cuk}
{{\'C}uk} M.,  {Dones} L.,   {Nesvorn{\'y}} D.,  2016, \mn@doi [\apj]
  {10.3847/0004-637X/820/2/97}, \href
  {http://adsabs.harvard.edu/abs/2016ApJ...820...97C} {820, 97}

\bibitem[\protect\citeauthoryear{{Dintrans} \& {Rieutord}}{{Dintrans} \&
  {Rieutord}}{2000}]{Dintrans}
{Dintrans} B.,  {Rieutord} M.,  2000, \aap, \href
  {http://adsabs.harvard.edu/abs/2000A%26A...354...86D} {354, 86}

\bibitem[\protect\citeauthoryear{{Dunford}, {Piazza}  \& {Thompson}}{{Dunford}
  et~al.}{2017}]{Dunford}
{Dunford} B.,  {Piazza} E.,   {Thompson} J.~R.,  2017, {Cassini: the grand
  finale}, \url{https://saturn.jpl.nasa.gov/}

\bibitem[\protect\citeauthoryear{Flattery}{Flattery}{1967}]{Flattery}
Flattery T.~W.,  1967, Hough functions.
University of Chicago. Department of the Geophysical Sciences

\bibitem[\protect\citeauthoryear{{French}, {Nicholson}, {McGhee-French},
  {Lonergan}, {Sepersky}, {Hedman}, {Marouf}  \& {Colwell}}{{French}
  et~al.}{2016}]{French}
{French} R.~G.,  {Nicholson} P.~D.,  {McGhee-French} C.~A.,  {Lonergan} K.,
  {Sepersky} T.,  {Hedman} M.~M.,  {Marouf} E.~A.,   {Colwell} J.~E.,  2016,
  \mn@doi [\icarus] {10.1016/j.icarus.2016.03.017}, \href
  {http://adsabs.harvard.edu/abs/2016Icar..274..131F} {274, 131}

\bibitem[\protect\citeauthoryear{{Fuller}}{{Fuller}}{2014}]{Fuller-2014}
{Fuller} J.,  2014, \mn@doi [\icarus] {10.1016/j.icarus.2014.08.006}, \href
  {http://adsabs.harvard.edu/abs/2014Icar..242..283F} {242, 283}

\bibitem[\protect\citeauthoryear{{Fuller}, {Luan}  \& {Quataert}}{{Fuller}
  et~al.}{2016}]{Fuller-2016}
{Fuller} J.,  {Luan} J.,   {Quataert} E.,  2016, \mn@doi [\mnras]
  {10.1093/mnras/stw609}, \href
  {http://adsabs.harvard.edu/abs/2016MNRAS.458.3867F} {458, 3867}

\bibitem[\protect\citeauthoryear{{Goldreich}}{{Goldreich}}{1965}]{Goldreich-1965}
{Goldreich} P.,  1965, \mn@doi [\mnras] {10.1093/mnras/130.3.159}, \href
  {http://adsabs.harvard.edu/abs/1965MNRAS.130..159G} {130, 159}

\bibitem[\protect\citeauthoryear{{Goldreich} \& {Nicholson}}{{Goldreich} \&
  {Nicholson}}{1977}]{Goldreich-Nicholson}
{Goldreich} P.,  {Nicholson} P.~D.,  1977, \mn@doi [\icarus]
  {10.1016/0019-1035(77)90163-4}, \href
  {http://adsabs.harvard.edu/abs/1977Icar...30..301G} {30, 301}

\bibitem[\protect\citeauthoryear{{Goldreich} \& {Nicholson}}{{Goldreich} \&
  {Nicholson}}{1989a}]{Goldreich-Nicholson-1989a}
{Goldreich} P.,  {Nicholson} P.~D.,  1989a, \mn@doi [\apj] {10.1086/167664},
  \href {http://adsabs.harvard.edu/abs/1989ApJ...342.1075G} {342, 1075}

\bibitem[\protect\citeauthoryear{{Goldreich} \& {Nicholson}}{{Goldreich} \&
  {Nicholson}}{1989b}]{Goldreich-Nicholson-1989b}
{Goldreich} P.,  {Nicholson} P.~D.,  1989b, \mn@doi [\apj] {10.1086/167665},
  \href {http://adsabs.harvard.edu/abs/1989ApJ...342.1079G} {342, 1079}

\bibitem[\protect\citeauthoryear{{Goldreich} \& {Tremaine}}{{Goldreich} \&
  {Tremaine}}{1978}]{Goldreich-Tremaine-1978}
{Goldreich} P.,  {Tremaine} S.,  1978, \mn@doi [\apj] {10.1086/156203}, \href
  {http://adsabs.harvard.edu/abs/1978ApJ...222..850G} {222, 850}

\bibitem[\protect\citeauthoryear{{Goldreich} \& {Tremaine}}{{Goldreich} \&
  {Tremaine}}{1982}]{Goldreich-Tremaine}
{Goldreich} P.,  {Tremaine} S.,  1982, \mn@doi [\araa]
  {10.1146/annurev.aa.20.090182.001341}, \href
  {http://adsabs.harvard.edu/abs/1982ARA%26A..20..249G} {20, 249}

\bibitem[\protect\citeauthoryear{{Goldreich} \& {Wu}}{{Goldreich} \&
  {Wu}}{1999}]{Goldreich-Wu}
{Goldreich} P.,  {Wu} Y.,  1999, \mn@doi [\apj] {10.1086/306705}, \href
  {http://adsabs.harvard.edu/abs/1999ApJ...511..904G} {511, 904}

\bibitem[\protect\citeauthoryear{{Goodman} \& {Lackner}}{{Goodman} \&
  {Lackner}}{2009}]{Goodman-Lackner}
{Goodman} J.,  {Lackner} C.,  2009, \mn@doi [\apj]
  {10.1088/0004-637X/696/2/2054}, \href
  {http://adsabs.harvard.edu/abs/2009ApJ...696.2054G} {696, 2054}

\bibitem[\protect\citeauthoryear{Greenspan}{Greenspan}{1958}]{Greenspan-1958}
Greenspan H.~P.,  1958, Journal of fluid Mechanics, 4, 330

\bibitem[\protect\citeauthoryear{Greenspan}{Greenspan}{1968}]{Greenspan}
Greenspan H.,  1968, The Theory of Rotating Fluids.
Cambridge Monographs on Mechanics, Cambridge University Press, \url
  {https://books.google.com/books?id=2R47AAAAIAAJ}

\bibitem[\protect\citeauthoryear{{Guenel}, {Mathis}  \& {Remus}}{{Guenel}
  et~al.}{2014}]{Guenel}
{Guenel} M.,  {Mathis} S.,   {Remus} F.,  2014, \mn@doi [\aap]
  {10.1051/0004-6361/201424010}, \href
  {http://adsabs.harvard.edu/abs/2014A%26A...566L...9G} {566, L9}

\bibitem[\protect\citeauthoryear{{Hansen}, {Kawaler}  \& {Trimble}}{{Hansen}
  et~al.}{2004}]{Hansen}
{Hansen} C.~J.,  {Kawaler} S.~D.,   {Trimble} V.,  2004, {Stellar interiors :
  physical principles, structure, and evolution}.
Springer

\bibitem[\protect\citeauthoryear{{Hedman} \& {Nicholson}}{{Hedman} \&
  {Nicholson}}{2013}]{Hedman-Nicholson}
{Hedman} M.~M.,  {Nicholson} P.~D.,  2013, \mn@doi [\aj]
  {10.1088/0004-6256/146/1/12}, \href
  {http://adsabs.harvard.edu/abs/2013AJ....146...12H} {146, 12}

\bibitem[\protect\citeauthoryear{{Hedman} \& {Nicholson}}{{Hedman} \&
  {Nicholson}}{2016}]{Hedman}
{Hedman} M.~M.,  {Nicholson} P.~D.,  2016, \mn@doi [\icarus]
  {10.1016/j.icarus.2016.01.007}, \href
  {http://adsabs.harvard.edu/abs/2016Icar..279..109H} {279, 109}

\bibitem[\protect\citeauthoryear{Hodges}{Hodges}{1967}]{Hodges}
Hodges R.,  1967, Journal of Geophysical Research, 72, 3455

\bibitem[\protect\citeauthoryear{Jeffreys}{Jeffreys}{1952}]{Jeffreys}
Jeffreys H.,  1952, The Earth.
Cambridge University Press

\bibitem[\protect\citeauthoryear{{Kerswell}}{{Kerswell}}{2002}]{Kerswell}
{Kerswell} R.~R.,  2002, \mn@doi [Annual Review of Fluid Mechanics]
  {10.1146/annurev.fluid.34.081701.171829}, \href
  {http://adsabs.harvard.edu/abs/2002AnRFM..34...83K} {34, 83}

\bibitem[\protect\citeauthoryear{{Lainey} et~al.,}{{Lainey}
  et~al.}{2012}]{Lainey-2012}
{Lainey} V.,  et~al., 2012, \mn@doi [\apj] {10.1088/0004-637X/752/1/14}, \href
  {http://adsabs.harvard.edu/abs/2012ApJ...752...14L} {752, 14}

\bibitem[\protect\citeauthoryear{{Lainey} et~al.,}{{Lainey}
  et~al.}{2017}]{Lainey}
{Lainey} V.,  et~al., 2017, \mn@doi [\icarus] {10.1016/j.icarus.2016.07.014},
  \href {http://adsabs.harvard.edu/abs/2017Icar..281..286L} {281, 286}

\bibitem[\protect\citeauthoryear{{Longuet-Higgins}}{{Longuet-Higgins}}{1968}]{Longuet-Higgins}
{Longuet-Higgins} M.~S.,  1968, \mn@doi [Philosophical Transactions of the
  Royal Society of London Series A] {10.1098/rsta.1968.0003}, \href
  {http://adsabs.harvard.edu/abs/1968RSPTA.262..511L} {262, 511}

\bibitem[\protect\citeauthoryear{{Luan} \& {Goldreich}}{{Luan} \&
  {Goldreich}}{2017}]{Luan}
{Luan} J.,  {Goldreich} P.,  2017, \mn@doi [\aj] {10.3847/1538-3881/153/1/17},
  \href {http://adsabs.harvard.edu/abs/2017AJ....153...17L} {153, 17}

\bibitem[\protect\citeauthoryear{{Maas}, {Benielli}, {Sommeria}  \&
  {Lam}}{{Maas} et~al.}{1997}]{Maas}
{Maas} L.~R.~M.,  {Benielli} D.,  {Sommeria} J.,   {Lam} F.-P.~A.,  1997,
  \mn@doi [\nat] {10.1038/41509}, \href
  {http://adsabs.harvard.edu/abs/1997Natur.388..557M} {388, 557}

\bibitem[\protect\citeauthoryear{{Marley} \& {Porco}}{{Marley} \&
  {Porco}}{1993}]{Marley}
{Marley} M.~S.,  {Porco} C.~C.,  1993, \mn@doi [\icarus]
  {10.1006/icar.1993.1189}, \href
  {http://adsabs.harvard.edu/abs/1993Icar..106..508M} {106, 508}

\bibitem[\protect\citeauthoryear{{Mathis}, {Neiner}  \& {Tran Minh}}{{Mathis}
  et~al.}{2014}]{Mathis}
{Mathis} S.,  {Neiner} C.,   {Tran Minh} N.,  2014, \mn@doi [\aap]
  {10.1051/0004-6361/201321830}, \href
  {http://adsabs.harvard.edu/abs/2014A%26A...565A..47M} {565, A47}

\bibitem[\protect\citeauthoryear{{Mathis}, {Auclair-Desrotour}, {Guenel},
  {Gallet}  \& {Le Poncin-Lafitte}}{{Mathis} et~al.}{2016}]{Mathis-2016}
{Mathis} S.,  {Auclair-Desrotour} P.,  {Guenel} M.,  {Gallet} F.,   {Le
  Poncin-Lafitte} C.,  2016, \mn@doi [\aap] {10.1051/0004-6361/201527545},
  \href {http://adsabs.harvard.edu/abs/2016A%26A...592A..33M} {592, A33}

\bibitem[\protect\citeauthoryear{{Murray} \& {Dermott}}{{Murray} \&
  {Dermott}}{1999}]{Murray-Dermott}
{Murray} C.~D.,  {Dermott} S.~F.,  1999, {Solar system dynamics}.
Cambridge University Press

\bibitem[\protect\citeauthoryear{{Nicholson}, {French}, {Hedman}, {Marouf}  \&
  {Colwell}}{{Nicholson} et~al.}{2014a}]{Nicholson-I}
{Nicholson} P.~D.,  {French} R.~G.,  {Hedman} M.~M.,  {Marouf} E.~A.,
  {Colwell} J.~E.,  2014a, \mn@doi [\icarus] {10.1016/j.icarus.2013.09.002},
  \href {http://adsabs.harvard.edu/abs/2014Icar..227..152N} {227, 152}

\bibitem[\protect\citeauthoryear{{Nicholson}, {French}, {McGhee-French},
  {Hedman}, {Marouf}, {Colwell}, {Lonergan}  \& {Sepersky}}{{Nicholson}
  et~al.}{2014b}]{Nicholson-II}
{Nicholson} P.~D.,  {French} R.~G.,  {McGhee-French} C.~A.,  {Hedman} M.~M.,
  {Marouf} E.~A.,  {Colwell} J.~E.,  {Lonergan} K.,   {Sepersky} T.,  2014b,
  \mn@doi [\icarus] {10.1016/j.icarus.2014.06.024}, \href
  {http://adsabs.harvard.edu/abs/2014Icar..241..373N} {241, 373}

\bibitem[\protect\citeauthoryear{{Ogilvie}}{{Ogilvie}}{2013}]{Ogilvie-2013}
{Ogilvie} G.~I.,  2013, \mn@doi [\mnras] {10.1093/mnras/sts362}, \href
  {http://adsabs.harvard.edu/abs/2013MNRAS.429..613O} {429, 613}

\bibitem[\protect\citeauthoryear{{Ogilvie} \& {Lesur}}{{Ogilvie} \&
  {Lesur}}{2012}]{Ogilvie-Lesur}
{Ogilvie} G.~I.,  {Lesur} G.,  2012, \mn@doi [\mnras]
  {10.1111/j.1365-2966.2012.20630.x}, \href
  {http://adsabs.harvard.edu/abs/2012MNRAS.422.1975O} {422, 1975}

\bibitem[\protect\citeauthoryear{{Ogilvie} \& {Lin}}{{Ogilvie} \&
  {Lin}}{2004}]{Ogilvie-Lin}
{Ogilvie} G.~I.,  {Lin} D.~N.~C.,  2004, \mn@doi [\apj] {10.1086/421454}, \href
  {http://adsabs.harvard.edu/abs/2004ApJ...610..477O} {610, 477}

\bibitem[\protect\citeauthoryear{{Peale}}{{Peale}}{1999}]{Peale}
{Peale} S.~J.,  1999, \mn@doi [\araa] {10.1146/annurev.astro.37.1.533}, \href
  {http://adsabs.harvard.edu/abs/1999ARA%26A..37..533P} {37, 533}

\bibitem[\protect\citeauthoryear{{Penev}, {Sasselov}, {Robinson}  \&
  {Demarque}}{{Penev} et~al.}{2007}]{Penev}
{Penev} K.,  {Sasselov} D.,  {Robinson} F.,   {Demarque} P.,  2007, \mn@doi
  [\apj] {10.1086/507937}, \href
  {http://adsabs.harvard.edu/abs/2007ApJ...655.1166P} {655, 1166}

\bibitem[\protect\citeauthoryear{Pierce}{Pierce}{1974}]{Pierce}
Pierce J.,  1974, Almost All about Waves.
MIT Press, \url {https://books.google.com/books?id=tG6bQgAACAAJ}

\bibitem[\protect\citeauthoryear{{Remus}, {Mathis}, {Zahn}  \&
  {Lainey}}{{Remus} et~al.}{2012}]{Remus}
{Remus} F.,  {Mathis} S.,  {Zahn} J.-P.,   {Lainey} V.,  2012, \mn@doi [\aap]
  {10.1051/0004-6361/201118595}, \href
  {http://adsabs.harvard.edu/abs/2012A%26A...541A.165R} {541, A165}

\bibitem[\protect\citeauthoryear{{Shoji} \& {Hussmann}}{{Shoji} \&
  {Hussmann}}{2017}]{Shoji}
{Shoji} D.,  {Hussmann} H.,  2017, \mn@doi [\aap]
  {10.1051/0004-6361/201630230}, \href
  {http://adsabs.harvard.edu/abs/2017A%26A...599L..10S} {599, L10}

\bibitem[\protect\citeauthoryear{{Spilker}, {Pilorz}, {Lane}, {Nelson},
  {Pollard}  \& {Russell}}{{Spilker} et~al.}{2004}]{Spilker}
{Spilker} L.~J.,  {Pilorz} S.,  {Lane} A.~L.,  {Nelson} R.~M.,  {Pollard} B.,
  {Russell} C.~T.,  2004, \mn@doi [\icarus] {10.1016/j.icarus.2004.05.016},
  \href {http://adsabs.harvard.edu/abs/2004Icar..171..372S} {171, 372}

\bibitem[\protect\citeauthoryear{{Stevenson}}{{Stevenson}}{1979}]{Stevenson}
{Stevenson} D.~J.,  1979, \mn@doi [Geophysical and Astrophysical Fluid
  Dynamics] {10.1080/03091927908242681}, \href
  {http://adsabs.harvard.edu/abs/1979GApFD..12..139S} {12, 139}

\bibitem[\protect\citeauthoryear{Urban \& Seidelmann}{Urban \&
  Seidelmann}{2012}]{Urban}
Urban S.,  Seidelmann P.,  2012, Explanatory Supplement to the Astronomical
  Almanac.
University Science Books, \url {https://books.google.com/books?id=c8pLLwEACAAJ}

\bibitem[\protect\citeauthoryear{{Weinberg}, {Arras}, {Quataert}  \&
  {Burkart}}{{Weinberg} et~al.}{2012}]{Weinberg}
{Weinberg} N.~N.,  {Arras} P.,  {Quataert} E.,   {Burkart} J.,  2012, \mn@doi
  [\apj] {10.1088/0004-637X/751/2/136}, \href
  {http://adsabs.harvard.edu/abs/2012ApJ...751..136W} {751, 136}

\bibitem[\protect\citeauthoryear{{Witte} \& {Savonije}}{{Witte} \&
  {Savonije}}{1999}]{Witte-Savonije}
{Witte} M.~G.,  {Savonije} G.~J.,  1999, \aap, \href
  {http://adsabs.harvard.edu/abs/1999A%26A...350..129W} {350, 129}

\bibitem[\protect\citeauthoryear{{Wu} \& {Goldreich}}{{Wu} \&
  {Goldreich}}{2001}]{Wu}
{Wu} Y.,  {Goldreich} P.,  2001, \mn@doi [\apj] {10.1086/318234}, \href
  {http://adsabs.harvard.edu/abs/2001ApJ...546..469W} {546, 469}

\bibitem[\protect\citeauthoryear{{Zahn}}{{Zahn}}{1966}]{Zahn-1966}
{Zahn} J.~P.,  1966, Annales d'Astrophysique, \href
  {http://adsabs.harvard.edu/abs/1966AnAp...29..489Z} {29, 489}

\bibitem[\protect\citeauthoryear{{Zahn}}{{Zahn}}{1989}]{Zahn-1989}
{Zahn} J.-P.,  1989, \aap, \href
  {http://adsabs.harvard.edu/abs/1989A%26A...220..112Z} {220, 112}

\makeatother
\end{thebibliography}




\appendix

\section{Linear damping for g modes}\label{sec:gamma-g-mode}

\subsection{Turbulent viscosity}\label{sec:gamma-turb}

\begin{figure}
\includegraphics[width=0.9\linewidth]{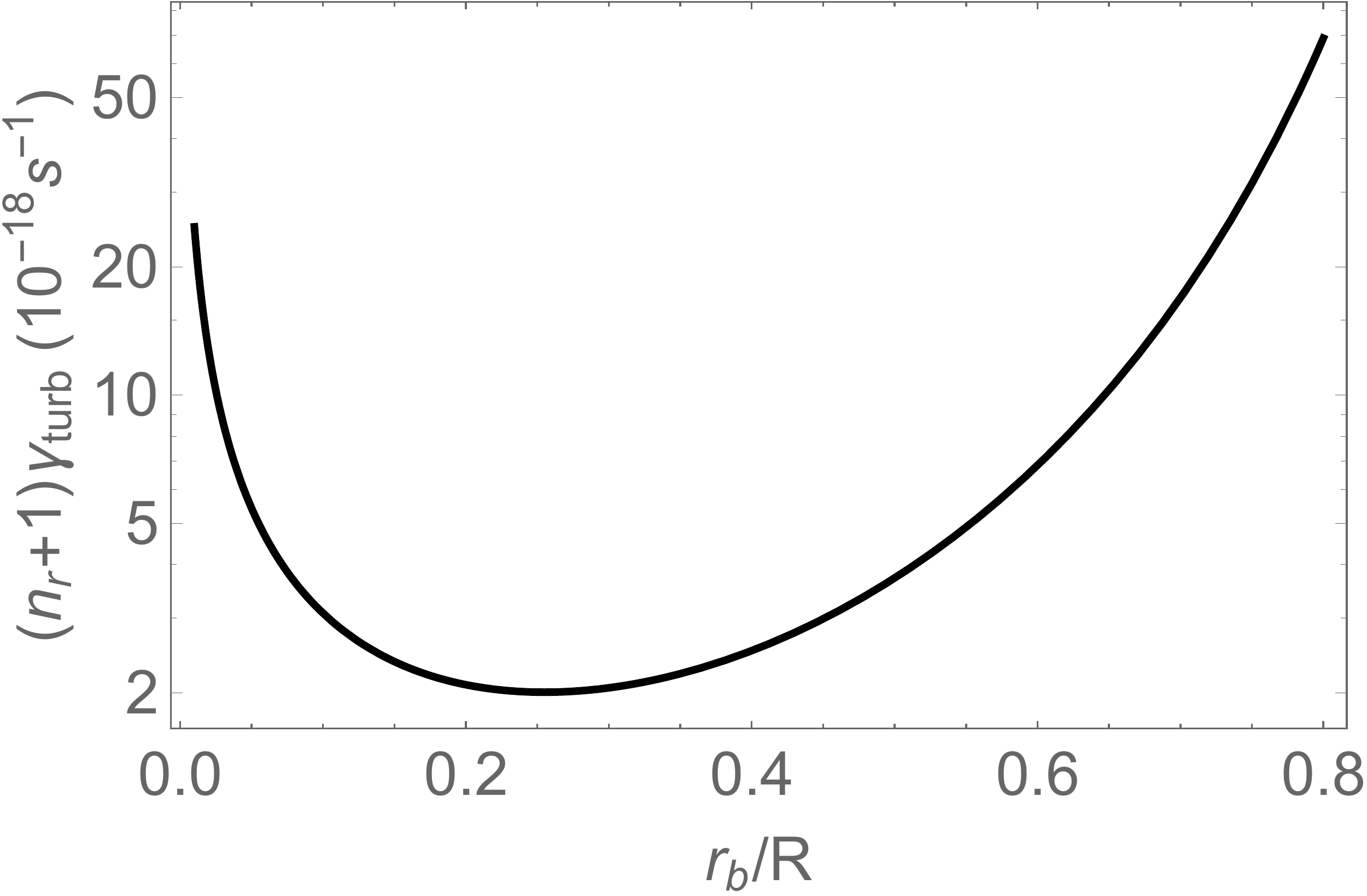}
\caption{\label{fig:gamma-rb}  Damping rate due to turbulent viscosity as a function of the radius of the bottom of the convection zone of Saturn.}
\end{figure}

G modes are evanescent in convection zones, and thus the length scale over which the displacement of the mode varies is $R-\rb$, where $r_b$ is the bottom of the convection zone. The energy dissipated per unit time in the convection zone is
\begin{equation}
\dot E_{\mathrm{turb}} \sim \int_{\rb}^{R} dr\, 4\pi r^2 \rho \nu_{\mathrm{turb}} \left(\dot\xi \over R-\rb\right)^2\, ,
\end{equation}
where the turbulent viscosity is $\nu_{\mathrm{turb}}\sim v_l l/3$, with $l$ the size of the convective eddy and $v_l$ its convective velocity. The turnover frequency, $v_l/l$, must be as fast as the oscillation for the turbulence to act like viscosity. However, the largest scale convective eddies, those as big as the local scale height, turn over very slowly compared to the oscillation frequency, and thus do not act like viscosity. Therefore, we need to consider sub-eddies in the turbulent cascade which we assume follows a Kolmogorov law, $v_l\propto l^{1/3}$. We adopt the prescription by \cite{Goldreich-Nicholson}.  Note that \cite{Zahn-1966} and \cite{Zahn-1989} propose a different prescription, and these two prescriptions have been under debate \citep{Penev,Ogilvie-Lesur}. Thus, setting
\begin{equation}\label{eq:Kolmogorov}
{v_l\over l}\sim {v_{\mathrm{cv}}\over H} \left(H\over l\right)^{2/3} \, ,
\end{equation}
to $\sigma$, we find the sub-eddies that act as viscosity generate
\begin{equation}
\nu_{\mathrm{turb}}\sim {1\over 3}{H v_{\mathrm{cv}}} \left({v_{\mathrm{cv}}\over H}{1\over\sigma}\right)^2\, ,
\end{equation}
Because $\sigma\gg v_{\mathrm{cv}}/H$, the turbulent viscosity is very small. 
The total energy of a mode with $n_r$ radial nodes is
\begin{equation}
E_{\mathrm{mode}}\sim (n_r+1)\int_{\rb}^R dr\, 4\pi r^2 \rho\, \dot\xi ^2\, ,
\end{equation}
assuming that the mode energy stored in the evanescent region is the same as that stored between each pair of consecutive radial nodes.
Because $\rho\, \dot\xi^2$ decreases outwards in the evanescent zone, the dissipation is dominated by the base of the evanescent zone, i.e.
\begin{equation}
\gamma_{\mathrm{turb}}\equiv {\dot E_{\mathrm{turb}} \over E_{\mathrm{mode}}}\sim {1\over (n_r+1)}{1\over (R-\rb)^2}\left(\left.v_{\mathrm{cv}}^3\over \sigma^2 H\right|_{\rb} \right)\, .
\end{equation}
We employ a polytrope of index unity \citep[e.g.][]{Hansen},
\begin{eqnarray}
z&=& A\, r\, ,\\
\rho(z) &=& \rho_c {\sin z\over z}\, ,
\end{eqnarray}
with
\begin{eqnarray}
A&=&{\pi \over R_S}\approx 5.4\times 10^{-10}\,\mathrm{cm^{-1}}\, ,\\
\rho_c&=& 2.253\,\mathrm{g\,cm^{-3}}\, ,
\end{eqnarray}
where $\rho_c$ is chosen to match Saturn's total mass.
Saturn's intrinsic luminosity, $L_{\mathrm{in}}\approx 8.45\times 10^{23}\,\mathrm{erg\, s^{-1}}$, is carried by convection, and it follows that
\begin{eqnarray}
4\pi r^2 \rho v_{\mathrm{cv}}^3 & =& L_{\mathrm{in}}\, .\label{eq:vcv-Lin}
\end{eqnarray}
The scale height is
\begin{eqnarray}
H &\sim & \left(-{d\ln p\over dr}\right)^{-1} =\left(-2{d\ln\rho\over dr}\right)^{-1} \,\\
&= & 2A\left(\cot z-{1\over z}\right)\, .
\end{eqnarray}
It follows that
\begin{eqnarray}
\gamma_{\mathrm{turb}}&\sim& (10^{-18}\sim 10^{ -17}) {1 \over (n_r+1)}\left(\sigma\over 2\OS \right)^{-2}\,\mathrm{s^{-1}}\, .\label{eq:gamma-turb}
\end{eqnarray}
Because $(n_r+1)\gamma$ appears together in the gravitational potential for a g mode (equation~\ref{eq:Phinl-final}), we plot $(n_r+1)\gamma_{\mathrm{turb}}$ in figure~\refnew{fig:gamma-rb}. 

Saturn spins quickly compared to its convective turnover frequency, i.e. $2\OS\gg \vcv/H$. Therefore, turbulence in Saturn is rotationally altered, since the Coriolis force makes it difficult to convect perpendicular to the spin axis. The major effect of rotation on convection is to make column-shaped eddies with their long axis along the spin axis. However, according to \cite{Barker}, rotation hardly affects turbulent viscosity. The derivation of the turbulent viscosity above depends on three conditions: 1, the radial size of eddy is $\sim H$; 2, a Kolmogorov law, equation~\refnew{eq:Kolmogorov}; 3, the relation between convective velocity and internal luminosity, equation~\refnew{eq:vcv-Lin}. All three conditions are still satisfied in the simple tank model of \cite{Barker}. Although rotationally modified convection in Saturn must be more complicated than the tank model, we speculate that rotation mainly modifies the shape of eddies, e.g., the ratio of their radial and horizontal length scales, rather than changing their absolute length scales. The absolute scale of eddies is primarily determined by the internal luminosity. We refer interested readers to \cite{Mathis-2016} for the modified turbulent viscosity applied on tidal flows in rotating turbulent convective layers. They combine the prescriptions for rotating convection by \cite{Stevenson} and \cite{Barker}. \cite{Stevenson} provide prescriptions in both the slow and fast rotating cases. \cite{Barker} confirms his results in the case of fast rotation, which is the case for Saturn. In the rapidly rotating regime the turbulent friction could be less efficient by several orders of magnitude when compared to the non-rotating case \citep{Mathis-2016}.

\subsection{Damping by heat diffusion\label{sec:gamma-diffusion}}
The local thermal timescale of Saturn is
\begin{equation}
\tau_{\mathrm{th}} = {4\pi r^2 p H\over L_{\mathrm{in}}} \sim 10^{17\sim 18}\,\mathrm{s}\, ,
\end{equation}
in the deep interior where g modes are likely to propagate. Diffusive processes scale inversely with the square of length, as revealed by the following general formula, 
\begin{eqnarray}
\rho c_p {\partial T\over \partial t}& =& -\bmath{\nabla}\cdot \bmath{F} \,\nonumber\\
&=& \bmath{\nabla}\cdot(k_{\mathrm{cond}}\bmath{\nabla}\, T)\, ,\label{eq:eqn-diff}
\end{eqnarray}
where $k_{\mathrm{cond}}$ is the effective conductivity. Therefore, the damping rate for g modes by diffusion is
\begin{eqnarray}
\gamma_{\mathrm{diff}}&\sim&{\int dr 4\pi r^2 \rho (\sigma\xi)^2 {(r k_r)^2\over \tau_{\mathrm{th,r}}}\over \int dr 4\pi r^2 \rho (\sigma\xi)^2}\, ,
\end{eqnarray}
where the integration is taken over the whole g mode cavity, and $\tau_{\mathrm{th,r}}$ is the thermal timescale at radius $r$. We observe that $\gamma_{\mathrm{diff}}$ is an average of the local diffusive timescale, $(r k_r)^2/\tau_{\mathrm{th,r}}$, weighted by the local kinetic energy of the mode. Because the local thermal timescale, $\tau_{\mathrm{th,r}}\sim p(r)H(r)/F(r)$, grows rapidly with depth, $\gamma_{\mathrm{diff}}$ is dominated by the first half wavelength. Therefore we have 
\begin{eqnarray}
\gamma_{\mathrm{diff}} &\sim & {\int_{\lambda_1} dr 4\pi r^2 \rho (\sigma\xi)^2 {(r k_r)^2\over \tau_{\mathrm{th,r}}}\over \int dr 4\pi r^2 \rho (\sigma\xi)^2}\, \nonumber\\
&\sim & \left.{(rk_r)^2\over \tau_{\mathrm{th}}}\right|_{r_b}\times  {\int_{\lambda_1} dr 4\pi r^2 \rho (\sigma\xi)^2 \over \int dr 4\pi r^2 \rho (\sigma\xi)^2}\,\nonumber\\
&\sim & \left.{(rk_r)^2\over \tau_{\mathrm{th}}}\right|_{r_b}\,{1\over (n_r+1)}\, ,
\end{eqnarray}
where in the last line we adopt the fact that the mode energy stored in the first half wavelength is about $1/(n_r+1)$ of the total mode energy. Employing that $k_r\sim (n_r+1)/r$, we derive that
\begin{eqnarray}
\gamma_{\mathrm{diff}}&\sim &{ (n_r+1)\over\tau_{\mathrm{th,b}}}\,\\
&\sim& (10^{-18}\sim 10^{-17}){(n_r+1)}\,\mathrm{s^{-1}}\, ,
\end{eqnarray}
which is similar to that due to turbulent viscosity. This derivation of $\gamma_{\mathrm{diff}}$ is a standard procedure, and thus we do not list details here. Interested readers are referred to e.g. \cite{Goldreich-Wu}. Here our intent is to justify that $\gamma_{\mathrm{diff}}$ in our case is almost independent of the specific diffusion mechanism, e.g., radiative diffusion, conductivity, etc. According to the second line of equation~(50) of \cite{Goldreich-Wu}\footnote{Entropy diffusion along horizontal direction is neglected, which is a good approximation for g mode, because $k_r\gg k_h$.}, 
\begin{equation}
\gamma_{\mathrm{diff}}\propto \int dr {\delta T\over T}{d\over dr}\left(\delta F\over F\right)\, ,
\end{equation}
where the integration is over the whole propagation cavity of a g mode which we assume coincides with the stably stratified region. The only term that the specific diffusion mechanism could affect is $\delta F/F$, because
\begin{equation}
F=-k_{\mathrm{cond}} {dT\over dr}\, ,
\end{equation}
so we obtain 
\begin{equation}
{\delta F\over F} = {\delta k_{\mathrm{cond}}\over k_{\mathrm{cond}}}+{\delta T\over T}-{d\xi_r\over dr} +\left(d\ln T\over dr\right)^{-1}{d\over dr}\left(\delta T\over T\right)\, ,\label{eq:dFoF}
\end{equation}
where $\delta k_{\mathrm{cond}}/k_{\mathrm{cond}}$ creates dependence on the specific diffusion mechanism. However, in the propagation cavity of a g mode, $k_r > 1/H$, where $1/H$ is the scale over which the unperturbed quantities vary. Therefore, terms containing a gradient of perturbed quantities, namely the last two terms in equation~\refnew{eq:dFoF}, dominate $\delta F/F$. Approximately,
\begin{equation}
{\delta F\over F} \approx -{d\xi_r\over dr} +\left(d\ln T\over dr\right)^{-1}{d\over dr}\left(\delta T\over T\right)\, .\label{eq:dFoF-approx}
\end{equation}
It follows that $\delta F/F$ is approximately independent of $k_{\mathrm{cond}}$, as is $\gamma_{\mathrm{diff}}$.

\section{Eulerian density perturbation}\label{sec:rho-prime}
Because we are interested in the first half wavelength, we can adopt a plane-parallel model with constant gravity, $g$.  Then equation (7) of \cite{Goldreich-Wu} applies, which we copy below,
\begin{equation}\label{eq:xiz}
\xi_z = {-g\sigma^2\over (g k_h)^2-\sigma^4}\left[{p\over\rho g}{d\over dz}\left(\delta p\over p\right)+\left(1-{k_h^2 p\over\sigma^2 \rho}\right)\left(\delta p\over p\right)\right]\, ,
\end{equation}
where $z$ is the vertical depth that increases downward, i.e., in the same direction of gravity, $g$. Here, $\delta$ denotes a Lagrangian perturbation. Note that this expression is general, without assuming $\nabla\cdot \mathbf{\xi}=0$ or $\delta\rho/\rho=0$.  We can neglect the $\sigma^4$ in the denominator because $\sigma^2\ll gk_h$ for a g mode in Saturn. For an adiabatic perturbation, $\delta\rho/\rho = \delta p/p/\Gamma_1$, and we have
\begin{eqnarray}
{\rho^\prime\over \rho}&=& {1\over\Gamma_1}{\delta p\over p}-\xi_z {d\ln\rho \over dz}\,\label{eq:rhop-1}\\
&\approx & \left({1\over\Gamma_1}-{d\ln\rho\over d\ln p}\right){\delta p\over p} +{\sigma^2\over g k_h}{d\ln\rho\over dz}{1\over k_h}{\delta p\over p}\,\\
&&+\left(d\ln\rho\over d\ln p\right){\sigma^2 \over gk_h}{1\over k_h} {d\over dz}\left(\delta p\over p\right)\, .
\end{eqnarray}
In the absence of a molecular weight gradient,
\begin{equation}
N^2= -g\left({1\over\Gamma_1}{d\ln p\over dz}-{d\ln\rho\over dz}\right)\, ,
\end{equation}
which may not be true but will not affect the generality of our discussion. Then we obtain
\begin{eqnarray}
{\rho^\prime \over \rho}&\approx & -{N^2 \over g}H_p\left(\delta p\over p\right)+{\sigma^2\over g k_h}{d\ln\rho \over dz}{1\over k_h}\left(\delta p\over p\right)\,\nonumber\\
&& +\left(d\ln\rho\over d\ln p\right){\sigma^2 \over gk_h}{1\over k_h} {d\over dz}\left(\delta p\over p\right)\, .\label{eq:rhop-2}
\end{eqnarray}
In the first half wavelength, $\lambda_1$, at the top of the propagation cavity we have the following approximations,
\begin{eqnarray}
k_z&\sim & k_h {N\over \sigma}\, ,\\
k_z &\sim & {1\over\lambda_1} \sim {1\over H}\, ,\\
H&\sim & H_p \sim H_\rho \, ,
\end{eqnarray}
where $H_\rho\equiv dz/d\ln\rho$ is the density scale height, and we do not distinguish it from the pressure scale height, $H_p\equiv dz/d\ln p = p/(\rho g)$, and we label both of them by $H$. Adopting these approximations, we realize that the three terms in equation~\refnew{eq:rhop-2} are of the same order of magnitude. For convenience, we choose the second term to represent $\rho^\prime/\rho$, i.e.,
\begin{equation}\label{eq:rhop-3}
{\rho^\prime\over \rho} \sim \left(\sigma^2\over gk_h\right){d\ln\rho \over dz} {1\over k_h} \left(\delta p\over p\right)\, .
\end{equation}
Next, let us relate $\rho^\prime/\rho$ to $\xi_z$. For Saturn, if we use a polytrope with index unity, we note that $k_h^2 p/(\rho\sigma^2)\gg 1$. In the first half wavelength, $p/(\rho g)=H_p\sim H$, $d(\delta p/p)/dz\sim -(\delta p/p)/\lambda_1$ and $H\sim\lambda_1$. It follows that the first two terms in $\xi_z$, equation~\refnew{eq:xiz}, are similar to each other and both are much smaller than the third term. Therefore, we obtain,
\begin{equation}\label{eq:xiz-app}
\xi_z \sim  H\left(\delta p\over p\right)\, .
\end{equation}
Combining equations~\refnew{eq:rhop-3} and \refnew{eq:xiz-app}, we obtain
\begin{equation}\label{eq:rhop-4}
\rho^\prime \sim {d\rho \over dz}\xi_z {\sigma^2\rho \over k_h^2 p} \sim {d\rho \over dr}\xi_r {\sigma^2\rho \over k_h^2 p}\, ,
\end{equation}
where we have changed the coordinate from $z$ to $r$.

\section{The first half wavelength dominates the gravity potential of g mode}\label{sec:first-domination}
We consider a simple Brunt-Vaisala frequency profile, i.e. $N=N_0$ between $r_c<r<r_b$. The number of radial nodes is
\begin{eqnarray}
\pi n_r &\sim & \int_{r_c}^{r_b}dr\, k_r \sim \int_{r_c}^{r_b}dr\, {K_n^{1/2}\over r}{N_0\over\sigma}\,\nonumber\\
&\sim  & K_n^{1/2}{N_0\over\sigma}\ln\left(r_b\over r_c\right)\, .
\end{eqnarray}
Similarly the $\tilde n$th node satisfies
\begin{equation}
\pi \tilde n \sim K_n^{1/2} {N_0\over \sigma}\ln\left(\rb\over r_{\tilde n}\right)\, .
\end{equation}
which means that the $\tilde n$th node is at a radius,
\begin{equation}
r_{\tilde n} = \rb \left(r_c\over r_b\right)^{\tilde n/n_r}\, ,\,\, (\tilde n= 0,\, 1,\, \dots,\, n_r-1) .
\end{equation}
The radial wavelength at the $\tilde n$th node is
\begin{equation}
\lambda_{\tilde n} \sim \left.{1\over k_r}\right|_{r_{\tilde n}}\sim \left({r\over K_n^{1/2}}{\sigma\over N_0}\right)_{r_{\tilde n}} \sim {r_{\tilde n}\over \pi n_r}\ln\left(r_b\over r_c\right)\, .
\end{equation}
The mode energy in each half wavelength is the same, and we express it in the $\tilde n$th half wavelength,
\begin{equation}
\Enode \sim  \left(r^2 \rho \lambda (\sigma \xi_h)^2\right)_{r_{\tilde n}}\sim \left(r^2 \lambda \rho N_0^2 \xi_r^2\right)_{r_{\tilde n}}\, ,
\end{equation}
yielding
\begin{equation}
\xi_{r,\tilde n} \sim (-1)^{\tilde n-1 } E_{\mathrm{node}}^{1/2}\left(\pi\, n_r \over \ln(r_b/r_c)\right)^{1/2}{1\over N_0}\left({1\over r^{3/2}}{1\over \rho^{1/2}}\right)_{r_{\tilde n}}\, .
\end{equation}
Note that $\xi_r$ switches sign in consecutive half wavelengths. We assume that $\xi_r$ is positive in the first half wavelength, which is an arbitrary assumption. Then it follows that
\begin{eqnarray}
\Sigma_{\tilde n}&\approx & \left.{d\rho\over dr}\right|_{\tilde n} \xi_{r,\tilde n}\, {\sigma^2\over g k_h}\lambda_{\tilde n}\,\\
 &\approx& (-1)^{\tilde n} E_{\mathrm{node}}^{1/2}{\sigma^2\over g k_h} \left(\ln(r_b/r_c)\over \pi\, n_r\right)^{1/2}{1\over N_0} \left({1\over r^{1/2}} {d\rho^{1/2}\over dr}\right)_{r_{\tilde n}}\, .\nonumber
\end{eqnarray}
According to equation~\refnew{eq:Phinl}, the $\tilde n$th half wavelength contributes a gravitational potential perturbation,
\begin{eqnarray}
\Phi_{\tilde n,\, nl}(r>r_{\tilde n})&=& -4\pi G\Sigma_{\tilde n} r_{\tilde n}{\mathcal{B}_{nl}\over (2l+1)}\left(r_{\tilde n}\over r\right)^{l+1}\,\label{eq:Phinl-node-line1} \\
&\sim & (-1)^{\tilde n-1}{2\pi^{1/2} \mathcal{B}_{nl}\over (2l+1)} G{E_{\mathrm{node}}^{1/2}\over N_0}\left(\ln(r_b/r_c)\over n_r\right)^{1/2}\,\nonumber\\
&&\times \left.{d\rho^{1/2}\over dr^{1/2}}\right|_{r_{\tilde n}}\,\left(r_{\tilde n}\over r\right)^{l+1}\left(\sigma^2\over g k_h\right)_{\tilde n}\, \label{eq:Phinl-node-line2}\\
&\sim & (-1)^{\tilde n-1}{2\pi \mathcal{B}_{nl}G E_{\mathrm{node}}^{1/2}\over (2l+1)K_n^{1/4}}{\sigma^{1/2}\over N_0^{3/2}}\left(r_b\over r \right)^{l+1}\,\nonumber\\
&&\times \left(d\rho^{1/2}\over dr^{1/2}\right)_{r_{\tilde n}}\left(r_c\over r_b\right)^{{\tilde n\over n_r}(l+1)}\left(\sigma^2\over g k_h\right)_{\tilde n} .\label{eq:Phinl-node-line3}
\end{eqnarray}
Note that $g k_h$ approaches a constant for $r_b/R_S\ll 1$. 
We consider two cases of $d\rho^{1/2}/dr^{1/2}$. First, let us assume it is constant. Then
\begin{eqnarray}
\Phi_{nl,\mathrm{tot}}&=&\sum_{\tilde n =0}^{n_r-1}\Phi_{\tilde n,\, nl} (r>R)\,\\
&\sim & \Phi_{\tilde n=0,\, nl}(r>R) \left( 1+\left(r_c\over r_b\right)^{(l+1)\over n_r}\right)^{-1}\, .
\end{eqnarray}
For a polytrope with index unity, $\rho=\rho_b \sin z/z$, where $z=\pi r/R$, and
\begin{equation}
{d\rho^{1/2}\over d r^{1/2}} = -{\rho_b^{1/2}\over 3}\left(\pi \over R\right)^2 r^{3/2}\, .
\end{equation}
It follows that
\begin{eqnarray}
\Phi_{nl,\mathrm{tot}}&=&\sum_{\tilde n =0}^{n_r-1}\Phi_{\tilde n,\, nl} (r>R)\,\\
&\sim & \Phi_{\tilde n=0,\, nl}(r>R) \left( 1+\left(r_c\over r_b\right)^{(l+5/2)\over n_r}\right)^{-1}\, .
\end{eqnarray}
In both cases, the outermost ($\tilde n=0$) half wavelength dominates the total $\Phi_{nl\,\mathrm{tot}}$, and the combined effect of all other half wavelengths is to reduce the contribution from the first half wavelength by a factor of two, at most.

\section{Potential of f modes exciting density waves in C ring}\label{sec:f-mode}
We assume the potential perturbation of an f mode outside Saturn to have spatial dependence
\begin{eqnarray}
\Phi_f &=&  \Delta\Phi_f \left(R_S\over r\right)^{l+1} \bar P_{lm}(\cos\theta_I) \cos(m\varphi_I -\sigma t)\, \\
&\sim & \Delta\Phi_f \left(R_S\over r\right)^{m+1} \cos(m\varphi_I-\sigma t)\, .
\end{eqnarray}
where $\theta_I$ and $\varphi_I$ are the colatitude and longitude measured in the inertial frame. We set $\theta_I=\pi/2$ because the C ring lies in the equatorial plane of Saturn. We consider the case $l=m$, which, for a given $m$, induces the strongest perturbation in the C ring. It excites a density wave at an outer Lindblad resonance \citep{Hedman-Nicholson}. Thus its pattern speed, i.e., the azimuthal phase speed, $\dot\varphi_I=\sigma/m$, satisfies
\begin{equation}
{\sigma\over m} = {m+1\over m} \Omega_{\mathrm{orb}}(r_L)\, ,
\end{equation} 
at the outer Lindblad resonance radius, $r_L$. Here $m$ is a positive integer, but note that \cite{Hedman-Nicholson} denote a negative azimuthal order for an outer Lindblad resonance. We express $r$ and $\varphi_I$ in terms of osculating elements of a ring particle's orbit to the first order of eccentricity \citep{Murray-Dermott},
\begin{eqnarray}
r&\approx & a(1-e\cos\lambda)\, ,\\
\varphi_I &\approx & \lambda+ \varpi +2 e\sin\lambda\, ,
\end{eqnarray}
where $\lambda$ is the mean longitude with revolution rate $\dot\lambda=n(a)$, $\varpi$ is the longitude of pericenter, $a$ is the semi-major axis, and $e$ is eccentricity. We submit these expressions into $\Phi_f$ and retain the term slowly varying with time, 
\begin{equation}
\Phi_{f,s} \sim {3m+1\over 2}\left(R\over a\right)^{m+1}\Delta\Phi_f e \cos\phi\, ,
\end{equation}
where $s$ denotes `slow', and $\phi\equiv (m+1)\lambda-\sigma t +m\varpi$. Since $\sigma = (m+1) n(r_L)$, and $\dot\lambda = \Omega_{\mathrm{orb}}(r)$, and $\varpi$ changes slowly with time, this term varies slowly with time near $r=r_L$ and it dominates the secular perturbation to the ring. This perturbing potential pumps eccentricity at a rate \citep{Goldreich-Tremaine},
\begin{eqnarray}
{de\over dt}&=& {1\over n a^2 e}{\partial \Phi_{f,s}\over \partial\varpi}\, \\
&\sim & -n {m(3m+1)\over 2} \left(R_S\over a\right)^m {\Delta\Phi_f\over (GM_S/R_S)}\sin\phi ,
\end{eqnarray}
which is maximized if $\phi=\pi/2$ or $3\pi/2$.

The dispersion relation for density waves in a self-gravitating disk in which pressure is negligible \citep{Goldreich-Tremaine-1978} is
\begin{equation}
(\sigma-m \Omega_{\mathrm{orb}}(r))^2 = \kappa(r)^2 -2\pi G \Sigma |k|\, ,
\end{equation}
where $\kappa(r)$ is the epicyclic frequency and is $\approx \Omega_{\mathrm{orb}}(r)$, $\Sigma$ is the mass surface density, and $k$ is the wave number. Near the outer Lindblad resonance, we have
\begin{equation}
|k| \sim {m+1\over G\Sigma} \left. {d n^2\over dr}\right|_{r_L}(r-r_L)\sim {(m+1)\over \Sigma}{M_S\over r_L^4}(r_L-r)\, ,
\end{equation}
which requires $r<r_L$, i.e., the density wave generated at the outer Lindblad resonance propagates inward. The first half wavelength is given by
\begin{equation}
\pi = \int_{r_L-\lambda_1}^{r_L}|k| dr\, ,
\end{equation}
yielding
\begin{equation}
\lambda_1 \sim {r_L\over (m+1)^{1/2}}\left(\Sigma r_L^2\over M_S\right)^{1/2}\, .
\end{equation}
The group velocity is 
\begin{equation}
v_g ={\partial \sigma\over\partial k} \sim -{\pi G\Sigma \over n}\, .
\end{equation}
The duration of the perturbing potential being in phase with the particle's epicyclic motion is 
\begin{equation}
t_{\mathrm{coh}} \sim {\lambda_1\over v_g} \sim {1\over n}{1\over \pi (m+1)^{1/2}}\left(M_S\over\Sigma r_L^2\right)^{1/2}\, .
\end{equation}
Therefore, the eccentricity changes by 
\begin{eqnarray}
\Delta e&\sim& {de\over dt} t_{\mathrm{coh}}\,\\
&\sim& {\Delta\Phi_f\over (GM_S/R_S)}\left(M\over\Sigma r_L^2\right)^{1/2}\left(R_S\over r_L\right)^m{m(3m+1)\over 2\pi (m+1)^{1/2}}\, .\nonumber\\
\end{eqnarray}

In order to generate a significant density variation, the excited epicyclic motion needs to cause orbits of different particles to almost cross. \cite{Borderies} derive the following criterion using a streamline model, i.e.,
\begin{equation}
q\sim \left(de\over d\ln a\right)^2\, .
\end{equation}
needs to be a good fraction of unity\footnote{There is a second term in the definition of $q$ in \cite{Borderies} describing the `twisting' effect as the longitude of pericenters change with $a$. This term is similar to the first term, $(de/d\ln a)^2$, in the first half wavelength, and therefore for our order-of-magnitude estimate we only estimate the first term.}. In our case,
\begin{equation}
q\sim \left(a {\Delta e\over \lambda_1}\right)^2 \, .
\end{equation}
Equating $q$ to unity leads to
\begin{equation}
\label{eqn:fmode}
{\Delta\Phi_f \over (GM_S/R_S)} \sim {2\pi \over m(3m+1)}\left(r_L\over R_S\right)^m \left(\Sigma\, r_L^2\over M_S\right)\, ,
\end{equation}
where we have replaced $a$ by the outer Lindblad radius, $r_L$. 
We calculate the value of equation \ref{eqn:fmode} using the density waves in the C ring reported by \cite{Hedman-Nicholson}, by adopting the surface density in their table 6 and the resonant location in their table 4. We list the corresponding potential perturbations due to f modes in table~\refnew{tab:f-mode}. We show that they are below the anticipated sensitivity of the gravity measurement of {\it Cassini}.


\bsp	
\label{lastpage}
\end{document}